\documentclass[english,aps,prd,a4paper,groupedaddress,preprintnumbers,floatfix,nofootinbib,showpacs,12pt]{revtex4}

\usepackage{graphicx}
\usepackage{enumerate}
\usepackage{amsmath}
\usepackage{amssymb}
\usepackage{amsfonts}
\usepackage{xcolor}
\usepackage{url}
\usepackage{hyperref}
\usepackage{booktabs}
\usepackage{caption}
\usepackage[utf8]{inputenc}

\usepackage{arydshln}
\usepackage{natbib}

\hypersetup{colorlinks=true,linkcolor=redLinks,citecolor=greenLinks,
urlcolor=redLinks,
pdfborder={0 0 1}}
\hypersetup{
    bookmarks=true,         % show bookmarks bar?
    unicode=false,          % non-Latin characters in Acrobat?s bookmarks
    pdftoolbar=true,        % show Acrobat?s toolbar?
    pdfmenubar=true,        % show Acrobat?s menu?
    pdffitwindow=false,     % window fit to page when opened
    pdfstartview={FitH},    % fits the width of the page to the window
    pdftitle={My title},    % title
    pdfauthor={Author},     % author
    pdfsubject={Subject},   % subject of the document
    pdfcreator={Creator},   % creator of the document
    pdfproducer={Producer}, % producer of the document
    pdfkeywords={keyword1, key2, key3}, % list of keywords
    pdfnewwindow=true,      % links in new PDF window
    colorlinks=false,       % false: boxed links; true: colored links
    linkcolor=red,          % color of internal links (change box color with linkbordercolor)
    citecolor=green,        % color of links to bibliography
    filecolor=magenta,      % color of file links
    urlcolor=cyan           % color of external links
}

\newcommand{\bi}{\begin{itemize}}
\newcommand{\ei}{\end{itemize}}

\begin{document}
\title{Interplay between nonstandard and nuclear constraints in 
  coherent elastic neutrino-nucleus scattering experiments.}
\author{B. C. Ca\~nas$^{1}$} \email{blanca.canas@unipamplona.edu.co}
\author{E. A. Garc\'es$^2$} \email{egarces@fisica.unam.mx}
\author{O. G. Miranda$^3$} \email{omr@fis.cinvestav.mx}
\author{A. Parada$^4$} \email{alexander.parada00@usc.edu.co}
\author{G. Sanchez Garcia$^3$} \email{gsanchez@fis.cinvestav.mx}
\affiliation{$^1$~Universidad de Pamplona, Km 1, v\'ia salida
  a Bucaramanga, Campus Universitario, 543050, Pamplona, Colombia}
 \affiliation{$^2$~Instituto~de F{\'{\i}}sica, 
 Universidad~Nacional Aut\'onoma de M\'exico, 
 Apdo. Postal 20-364, CDMX 01000, M\'exico.} 
\affiliation{$^3$~Departamento de F\'isica, Centro de Investigaci\'on
  y de Estudios Avanzados del IPN, Apdo. Postal 14-740, 07000 Ciudad
  de M\'exico, M\'exico.}    
  \affiliation{$^4$~Facultad de Ciencias B\'asicas, Universidad Santiago de
  Cali, Campus Pampalinda, Calle 5 No. 6200, 760035, Santiago de Cali,
  Colombia.} 

\begin{abstract}\noindent
New measurements of the coherent elastic neutrino-nucleus scattering
(CEvNS) are expected to be achieved in the near future by using two
neutrino production channels with different energy distributions: the
very low energy electron antineutrinos from reactor sources and the
muon and electron neutrinos from spallation neutron sources (SNS)
with a relatively higher energy. Although precise measurements of this
reaction would allow for an improved knowledge of standard and beyond the
Standard Model physics, it is important to distinguish the different
new contributions to the process.  We illustrate this idea by
constraining the average neutron root mean square (rms) radius of the
scattering material, as a standard physics parameter, together with
the nonstandard interactions (NSI) contribution as the new physics
formalism.  We show that the combination of experiments with different
neutrino energy ranges could give place to more robust constraints on
these parameters as long as the systematic errors are under control.
\end{abstract}

\maketitle

\section{Introduction}
\label{sec:1}

Four decades after its theoretical prediction~\cite{Freedman:1973yd},
the COHERENT Collaboration~\cite{Akimov:2017ade} has eventually
accomplished the challenge of first observing the coherent elastic
neutrino-nucleus scattering (CEvNS) phenomenon. This measurement was
achieved with neutrinos coming from a spallation neutron source (SNS)
with energies up to 52.8 MeV.  The importance of this neutrino
reaction lays on the fact that it can be used as a powerful tool for
precision measurements at low energy ranges, as well as a mechanism
for ruling out or confirming a variety of new physics
scenarios~\cite{Papoulias:2019txv,Khan:2019cvi,Coloma:2019mbs,Papoulias:2019xaw,AristizabalSierra:2019ykk,Miranda:2019skf,Altmannshofer:2018xyo,Cadeddu:2018dux,Canas:2018rng,Billard:2018jnl,Denton:2018xmq,Farzan:2018gtr,Kosmas:2017tsq,Cadeddu:2017etk,Canas:2017umu,Ge:2017mcq,Garces:2017qkm,Shoemaker:2017lzs,Kosmas:2017zbh,Dutta:2015nlo,Kosmas:2015vsa,Kosmas:2015sqa,Barranco:2011wx,Barranco:2007tz}.

In addition, future dark matter experiments are nearing their sensitivity to
the neutrino floor~\cite{Billard:2014yka}, and an accurate
characterization of the detector target materials through
CEvNS~\cite{Boehm:2018sux}, using diverse neutrino data from SNS and
reactors, will be important in discriminating the neutrino background
from a true wimp signature~\cite{Papoulias:2018uzy,AristizabalSierra:2017joc}.

The CEvNS reaction in the energy regime of the SNS depends on nuclear
form factors, introducing their 
own source of
errors~\cite{Papoulias:2018uzy,AristizabalSierra:2017joc}, 
which
unfortunately tend to be difficult to have under control. In contrast, the
detection of CEvNS from reactor neutrinos (with an average energy of
$<E_\nu> \approx 4$~MeV) in the foreseeable future, would allow the
study of neutrino physics in the very low energy window, a regime
where there is nearly no contribution of
the nuclear form factor.  Many experiments under commission
are in the quest for the first detection of CEvNS using
reactor neutrinos as a source; among them we have:
TEXONO~\cite{Wong:2005vg}, CONUS~\cite{Lindner:2016wff},
NU-CLEUS~\cite{Rothe:2018ulo},
CONNIE~\cite{Aguilar-Arevalo:2016qen,Aguilar-Arevalo:2016khx,Chavez-Estrada:2017gni,Aguilar-Arevalo:2019jlr,
  Aguilar-Arevalo:2019zme}, MINER~\cite{Agnolet:2016zir},
RED100~\cite{Akimov:2012aya}, and RICOCHET~\cite{Billard:2016giu}.

Because of their different characteristics, it can be expected that
the combination of CEvNS from reactor and SNS fluxes can accurately
constrain standard and nonstandard physics. It would be desirable to
investigate how the correlation between different parameters
describing standard and nonstandard interactions can be disentangled
combining various experimental setups. The interplay among different
observables has been analyzed in previous studies, including the
correlation between the weak mixing angle and the neutron
skin~\cite{Huang:2019ene}, as well as between the form factor
uncertainties and new physics constraints in current and future SNS
CEvNS experiments~\cite{AristizabalSierra:2019zmy}.  On the other
hand, in this work we simultaneously study the potential to measure
the relevant parameter for the nuclear form factor (the neutron rms
radius $R_n$) and the restriction to new physics in the nonstandard
interactions (NSI) formalism, a model independent picture able to
describe many beyond Standard Model scenarios for neutrino
interactions~\cite{Farzan:2017xzy,Miranda:2015dra,Ohlsson:2012kf,Dev:2019anc}.
Until now, NSI have been extensively studied in the context of
CEvNS~\cite{Barranco:2005yy,Barranco:2007tz,Barranco:2011wx,Billard:2018jnl,Kosmas:2017tsq,Esteban:2018ppq,Denton:2018xmq,AristizabalSierra:2018eqm}
and its current constraints are already useful for global
analysis~\cite{deSalas:2017kay,globalfit,Esteban:2016qun,Capozzi:2018ubv},
specially to break some of the well-known degeneracy problems leading
to the LMA-dark solution~\cite{miranda:2004nb,Coloma:2016gei} and to
the possible degeneracy in probing CP violation in neutrino
oscillations~\cite{Miranda:2016wdr,Forero:2016cmb}.

\section{Nonstandard Neutrino Interactions and CEvNS}

As we mentioned above, the detection of CEvNS has been pursued for a
long time, and since the first results were presented by the COHERENT
Collaboration~\cite{Akimov:2017ade}, the interest to confirm such
reaction has increased. In this section we briefly introduce the main
characteristics of the CEvNS process, to be used in later
calculations. Within the SM, the CEvNS differential cross section is
given
by~\cite{Drukier:1983gj,Patton:2012jr,Papoulias:2015vxa,Bednyakov:2018mjd}
\begin{equation} 
\left(\frac{d\sigma}{dT}\right)_{\rm SM}^{\rm coh} = \frac{G_{F}^{2}M}{\pi}\left[1-\frac{MT}{2E_{\nu}^{2}}\right]
 [Zg^{p}_{V}F_Z(q^2)+Ng^{n}_{V}F_N(q^2)]^{2}. \label{eq:00}
\end{equation}
where, $M$ is the mass of the nucleus, $E_{\nu}$ the incoming neutrino energy, $T$ is the nucleus recoil energy, 
$F_{Z,N}(q^2)$ are the nuclear form factors, and
the neutral current vector couplings (including
radiative corrections) are given by~\cite{Barranco:2005yy}:
\begin{eqnarray}
\nonumber g_{V}^{p} &=&\rho_{\nu
  N}^{NC}\left(\frac{1}{2}-2\hat{\kappa}_{\nu
  N}\hat{s}_{Z}^{2}\right)+2\lambda^{uL}+2\lambda^{uR}+\lambda^{dL}+\lambda^{dR}\\ g_{V}^{n}
&=&-\frac{1}{2}\rho_{\nu
  N}^{NC}+\lambda^{uL}+\lambda^{uR}+2\lambda^{dL}+2\lambda^{dR}
\end{eqnarray}
\noindent where $\rho_{\nu N}^{NC}=1.0086$,
$\hat{s}_{Z}^{2}=\sin^2\theta_{W}=0.2312$, $\hat{\kappa}_{\nu
  N}=0.9978$, $\lambda^{uL}=-0.0031$, $\lambda^{dL}=-0.0025$, and
$\lambda^{dR}=2\lambda^{uR}=7.5 \times 10^{-5}$~\cite{Patrignani:2016xqp}. 
Notice that the previous relations depend on fundamental parameters such as 
the weak mixing angle, $\sin\theta_W$, on nuclear physics through the 
form factors $F_Z(q^2)$ and $F_N(q^2)$, and on the specific detection target through the proportion of protons to neutrons $Z/N$. 

For the purposes of this work, we are also interested in the search
for new physics. As mentioned before, a common framework is that of
nonstandard neutrino
interactions~\cite{Ohlsson:2012kf,Miranda:2015dra,Farzan:2017xzy}. In
this scenario, new terms containing nonuniversal and flavor changing
currents are present. These terms are parametrized as dimensionless
coefficients $\varepsilon _{\alpha \beta}^{qV}$ (with $ q = u, d, V =
L, R$ and $\alpha, \beta = e, \mu, \tau$.) proportional to the Fermi
constant. The parameters for which $\alpha = \beta$ refer to
nonuniversal interactions, while those with $\alpha \neq \beta$
correspond to flavor-changing terms. By introducing these parameters,
the CEvNS cross section in the spinless limit, for $T << E_{\nu}$, is
given
by~\cite{Barranco:2005yy,Scholberg:2005qs,AristizabalSierra:2018eqm,
  Dent:2017mpr,Lindner:2016wff,Coloma:2017egw}:
\begin{equation}
\begin{aligned}
\frac{\mathrm{d} \sigma }{\mathrm{d} T}\left ( E_{\nu },T \right ) \simeq& \frac{G_{F}^{2}M}{\pi }\left ( 1-\frac{MT}{2E_{\nu }^{2}} \right )\Big\{ \left [ Z\left ( g_{V}^{p}+2\varepsilon _{ee}^{uV}+\varepsilon _{ee}^{dV} \right )F_{Z}^{V}(Q^{2})+N\left ( g_{V}^{n}+\varepsilon _{ee}^{uV}+2\varepsilon _{ee}^{dV}  \right ) F_{N}^{V}(Q^{2})\right ]^{2}\\
&+\sum_{\alpha }\left [ Z\left ( 2\varepsilon _{\alpha e}^{uV}+\varepsilon _{\alpha e}^{dV} \right )F_{Z}^{V}(Q^{2})+N\left ( \varepsilon _{\alpha e}^{uV}+2\varepsilon _{\alpha e}^{dV} \right )F_{N}^{V}(Q^{2}) \right ]^{2} \Big\} ,
\end{aligned}
\label{CrossN}
\end{equation}
in the expression above, we have exemplified NSI for the case of
an incoming electron (anti)neutrino flux, as is the case of reactor
neutrino experiments.  The corresponding cross section for an incident
muon and tau (anti)neutrino is straightforward to obtain by replacing
$e \leftrightarrow \mu, \tau$, respectively.  Comparing this last
equation with the Standard Model case of Eq.~(\ref{eq:00}), we can
notice that, besides the dependence on the previous parameters, such
as the nuclear form factors, we now have the dependence on the NSI
parameters $\varepsilon^{qV}_{\alpha\beta}$. A first approach has been
to study each of the observables separately.  With this method it has
been possible to obtain constraints for the neutron rms
radius~\cite{Cadeddu:2017etk,AristizabalSierra:2019zmy,Papoulias:2019lfi},
NSI parameters~\cite{Kosmas:2017tsq,Coloma:2017ncl}, and neutrino
electromagnetic properties~\cite{Miranda:2019wdy, Kosmas:2017tsq,
  Parada:2019gvy}, among others. Most of these analyses have been done
for one parameter at a time. As a next step in this direction, we aim
to forecast whether a combination of different CEvNS measurements
disentangles information from different observables, in order to have
more reliable constraints.

In the next sections we will discuss this case of NSI as well as the
neutron radius distribution, the latter being a parameter contained
within the nuclear form factor.

\section{ CEvNS experiments with reactor antineutrinos}
\label{sec:3}

Reactors have been a useful source of antineutrinos since the first
neutrino signal  ever detected~\cite{Reines:1956rs,Cowan:1992xc}. The
most used reaction to detect them has been the inverse beta
decay~\cite{An:2012eh,Ahn:2012nd,Abe:2011fz,Apollonio:1999ae} while
neutrino-electron scattering has given complementary information,
although with less statistics due to the smaller cross
section~\cite{Deniz:2009mu,Reines:1976pv}. A measurement of CEvNS from
reactor antineutrinos will provide yet another channel to measure the
antineutrino flux and could give complementary information to
constrain, for instance, a sterile neutrino signal~\cite{Dutta:2015nlo,Canas:2017umu,Kosmas:2017zbh}.

The expected number of events to be detected in a CEvNS reactor
antineutrino experiment is computed by using the integral of the 
incoming neutrino flux times the cross section, 

\begin{equation}\label{n_events}
 N^{\rm}_{\rm events}=t\phi_{0}\frac{M_{\text{detector}}}{M}\int_{E_{\nu \rm min}}^{E_{\nu \rm max}}
 \lambda(E_{\nu})dE_{\nu}\int_{T_{\rm min}}^{T_{\rm max}(E_{\nu})}\left(\frac{d\sigma}
 {dT}\right)^{\rm coh} dT , 
\end{equation}
where $t$ is the exposure time of the experiment (that we will consider as 
one year), $\phi_0$ is the antineutrino flux from the reactor, $M_\text{detector}$ is the detector mass, $M$ is the nucleon mass, and $\lambda(E_\nu)$ is the antineutrino energy spectrum. 
Notice that in general, the cross section $d\sigma / 
 dT$ refers to either the Standard Model or to the
NSI one, which are given, respectively, by Eqs.~(\ref{eq:00})
and~(\ref{CrossN}). 

Due to the low energy antineutrino spectrum, the
momentum transfer is small and the nuclear form factors
are expected to be close to unity and almost constant. 
Therefore, the dependence on nuclear physics effects
will be rather weak for this kind of experiment.

For the antineutrino energy spectrum $\lambda(E_{\nu})$, we will use
the theoretical one discussed in
Refs.~\cite{Mueller:2011nm,Huber:2011wv}, which is parametrized as: 
\begin{equation}
\lambda(E_\nu) = \Sigma_l f_l \lambda_l(E_\nu) = \Sigma_l f_l \exp\big[ \Sigma_{k=1}^6 
\alpha_{kl}E_\nu^{k-1}\big], 
\label{eq:nuFlux}
\end{equation}
$f_l$ represents the fission fraction for the given isotope
($^{235}$~U, $^{239}$~Pu, $^{241}$~Pu, $^{238}$~U) in the reactor. For
the case of CONNIE, we will
consider~\cite{Aguilar-Arevalo:2016qen,Aguilar-Arevalo:2016khx}
(0.55:0.32:0.06:0.07,) while for CONUS, we use the fission
fractions~\cite{Lindner:2016wff} (0.58:0.30:0.05:0.07). For the
specific values of the coefficients $\alpha_{kl}$ we follow the
prescription given in~\cite{Mueller:2011nm}. This expression is valid
for energies above $2$~MeV, while for lower energies we use the values
reported in~\cite{Kopeikin:1997ve}.

With this information, we can make a forecast of the
sensitivity to NSI for a future CEvNS reactor antineutrino measurement. We
can consider that the experiment will measure the Standard Model (SM) predicted number
of events, $N^{exp}=N^{\rm SM}_{\rm events}$, with a given statistical
and systematic error. We can then compute a $\chi^2$ statistical
analysis as
\begin{equation}
  \label{Eq:chi}
  \chi^2 = \frac { (N^{\rm SM}_{\rm events}
-N^{th})^2 } {\sigma_{stat}^2 + \sigma_{syst}^2} .
\end{equation}

There are several experiments that are trying to observe CEvNS from a
reactor antineutrino flux.  In this work we will focus on two
particular experimental setups, the CONUS and the CONNIE proposals,
for which the main characteristics are listed in Table I.  One of the main
reasons for considering these experiments is that they use two
different materials (germanium and silicon, respectively) and
detection technologies, this will lead to complementary results for
the NSI constraints, specially because they have a different
proportion of protons to neutrons, a quantity to which the cross
section is sensitive.

\begin{table}
  \begin{tabular}{l c c c c c c c c} \hline \hline
& $T_{thres}$ & Baseline & $Z/N$  & Det. Tec. & Fid. Mass  \\ \hline \hline
CONNIE~\cite{Aguilar-Arevalo:2016qen,Aguilar-Arevalo:2016khx} 
& $28$~eV &30~m& $1.0$ & CCD (Si) & $0.1 -  1$~kg\\ 
CONUS~\cite{Lindner:2016wff}  &    
 100~eV & 10~m & $0.79$  & HPGe & $4 - 100$~kg \\ 
\hline \hline 
\end{tabular}\caption{\label{Tab:01} 
List of the experimental proposals to detect CEvNS in reactor
neutrinos. We show the information for the expected electron recoil
energy threshold (T thres ), the distance from the source to the
detector (Baseline) the detector technology, (Det. Tec), and the
fiducial mass (Fid. Mass).}
\end{table}

\begin{figure}[h] 
\begin{center}
\includegraphics[width=0.49\textwidth]{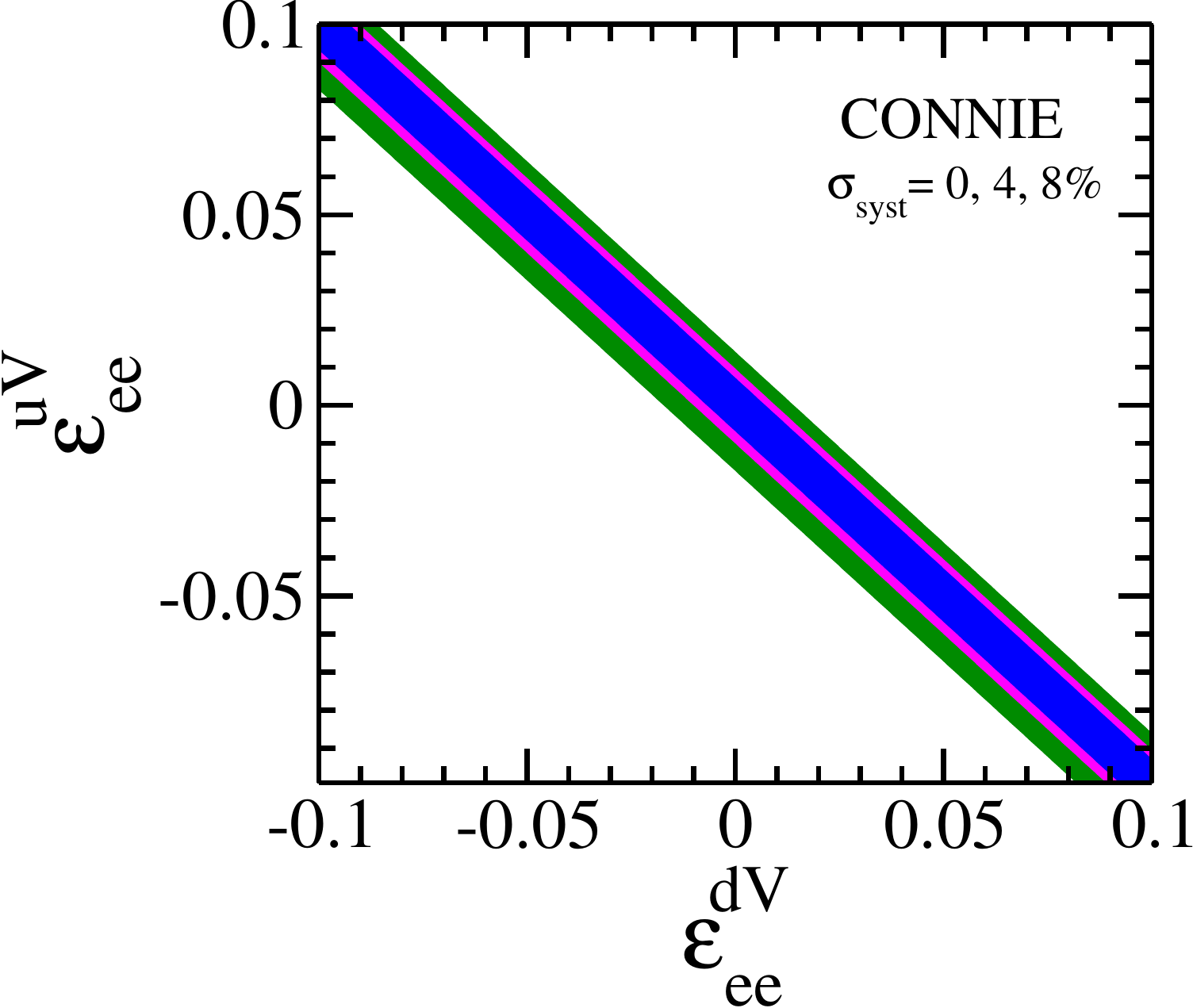}
\includegraphics[width=0.49\textwidth]{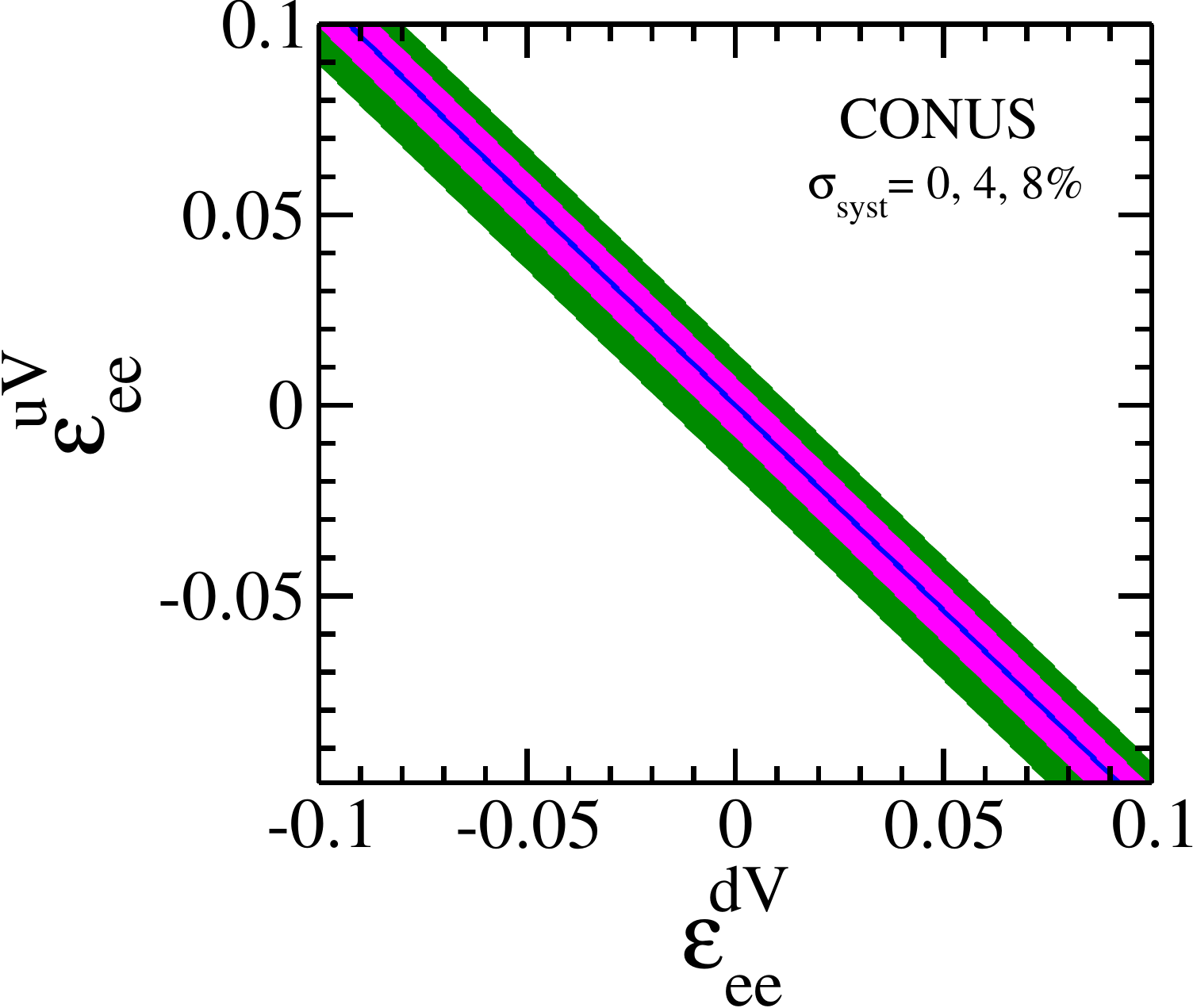}
\end{center}
\caption{Expected exclusion for the non universal NSI parameters in the case of detection 
of CEvNS at the reactor neutrino experiments CONNIE and CONUS, the colored regions
 indicate the exclusion with an overall systematic error of 0,  4 and 8\%.  \label{fig:reactorNU}}
\end{figure}

\begin{figure}[h] 
\begin{center}
\includegraphics[width=0.49\textwidth]{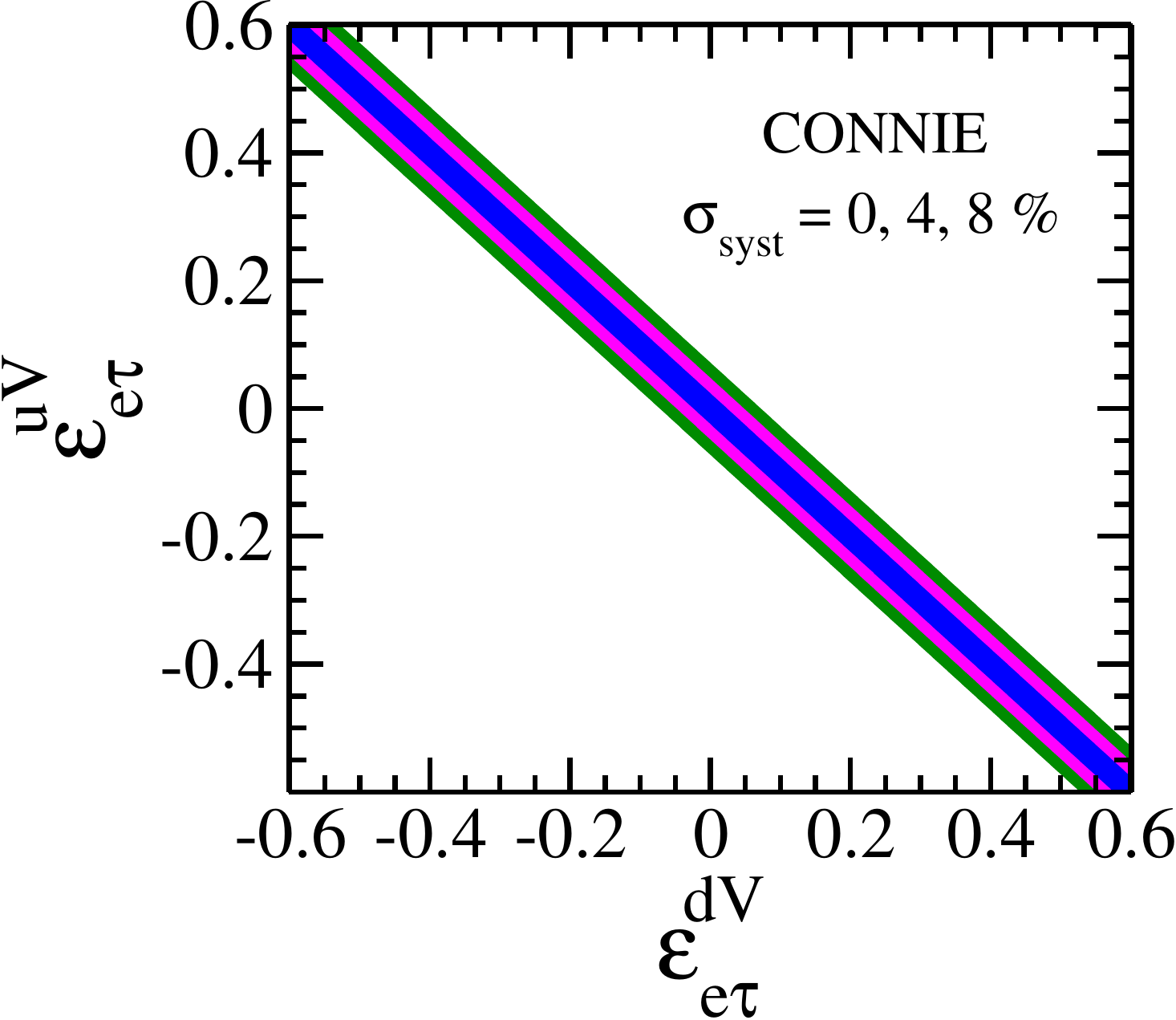}
\includegraphics[width=0.49\textwidth]{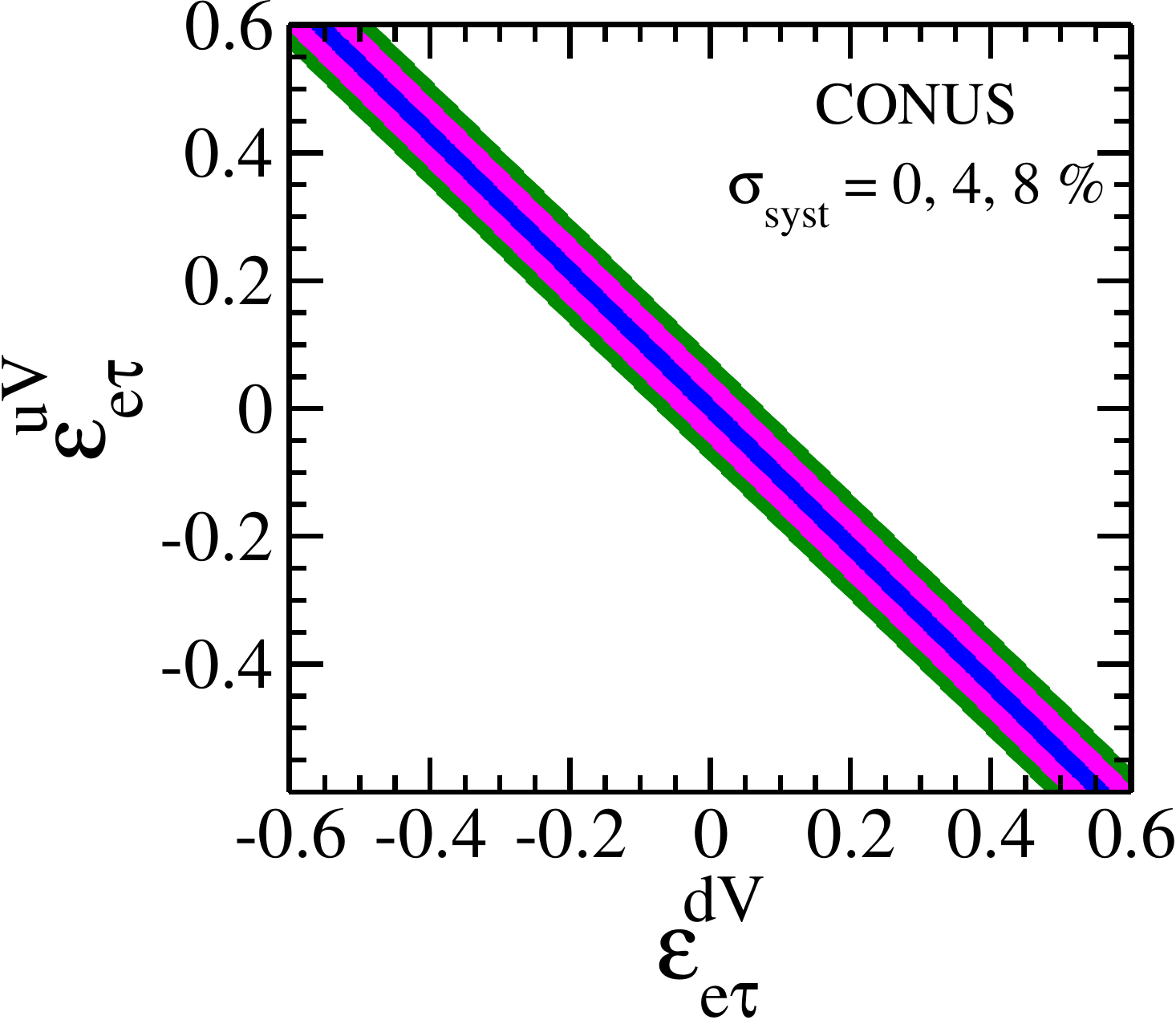}
\end{center}
\caption{Expected exclusion for the flavor changing NSI parameters  in the case of detection of CEvNS 
at the reactor neutrino experiments CONNIE and CONUS, the colored regions indicate the exclusion 
with an overall systematic error of 0,  4 and 8\%.  \label{fig:reactorFC}}
\end{figure}

After performing a $\chi^2$ analysis, we have found the corresponding
constraints on NSI for these two proposals. In each analysis we take
only two nonzero NSI parameters.  We include the statistical
uncertainty as $\sigma_{stat}=\sqrt{N^{\rm SM}_{\rm events}}$, with
$N^{\rm SM}_{\rm events}$ the expected number of events to be measured
in one year.  As for the systematics, it is difficult to forecast what
would be the final experimental error. However, for the case of
CONUS~\cite{Lindner:2016wff}, it has been considered that a $5$~\%
normalization error and a $3$~\% of other overall systematic
errors can be a realistic scenario. To be more conservative, in this
work, we will assume an $8$~\% contribution for the systematic errors,
considering that systematic uncertainties in quantities such as the
quenching factor (QF) can also play a significant role. For
completeness, we also computed the optimistic case of a $4$~\%
systematic error and, finally, the ideal case for which only
statistical uncertainties are taken into account; although it is
unexpected that the systematics could be smaller than the statistical
error, it gives an idea of the more optimistic achievable
constraint. The results are shown in Fig.~\ref{fig:reactorNU} for the
non-universal NSI parameters and in Fig.~\ref{fig:reactorFC} for the
flavor-changing ones. We can see that both proposals could give a good
restriction for the NSI parameters if they are considered one at a
time.  For the flavor-changing analysis we have only considered the
$e\tau$ channel, since the $e\mu$ NSI parameter has already been
constrained with other neutrino
processes~\cite{Farzan:2017xzy,Miranda:2015dra,Ohlsson:2012kf}. In
particular, we can see that, when we take one parameter at a time, the
constraint is quite strong and would improve current limits given by
other
experiments~\cite{Farzan:2017xzy,Miranda:2015dra,Ohlsson:2012kf}.
This information is of relevance for the future exact determination of
the standard oscillation parameters by long-baseline experiments, in
particular of the CP violating phase, $\delta_{CP}$, that could have a
different value if flavor changing NSI are
present~\cite{Forero:2016cmb}.

Moreover, if we combine the two results it might be possible to obtain a
more robust constraint in the sense that the parameter degeneracy can
be resolved. In this case, it is important to include the
systematic errors of the reactor antineutrino flux and their
correlations. We will see that the influence of such correlated error is 
mild and perhaps there could be other systematic uncertainties that would be 
more relevant, such as the QF. 

To illustrate the effect of the correlations for the reactor
antineutrino spectrum, we have considered it for the moment as the only
source of systematic uncertainty and made a complete $\chi^2$
analysis. For this analysis we closely follow the approach
described in Ref.~\cite{Huber:2004xh} where the covariance matrix is
diagonalized in order to simplify the numerical analysis. In this
framework we start with the covariance matrix
\begin{equation}
V_{kk'}^l = \delta \alpha_{kl}\delta\alpha_{k'l}\rho_{kk'}^l
\end{equation}
where $\delta\alpha_{kl}$ are the errors of the coefficients
$\alpha_{kk'}$ that appear in Eq.~(\ref{eq:nuFlux}) and
$\rho_{kk'}^l$ the corresponding correlation matrix (these values were 
reported in Ref.~\cite{Mueller:2011nm}). The systematic
error arising from the reactor antineutrino flux is given then by 
\begin{equation}
(\delta N_{l}^{\nu})^2 = \sum_{kk'} 
\frac{\partial  N^\nu_{l}}{\partial \alpha_{kl}} 
\frac{\partial  N^\nu_{l}}{\partial \alpha_{k'l}}
V_{kk'}^l.
\end{equation}
To work with the diagonal form of the covariance matrix, it is necessary to 
introduce new coefficients~\cite{Huber:2004xh}, $c_{kl}$, that are defined by 
the relation:  
\begin{equation}
\alpha_{kl} = \sum_{k'} \mathcal{O}^{l}_{k'k}c_{k'l} ,
\end{equation}
with $\mathcal{O}^{l}$ a rotation matrix such that it diagonalizes the covariance matrix $V^{l}$
\begin{equation}
\mathcal{O}^{l}
V^{l}
(\mathcal{O}^{l})^{T} = diag[(\delta c_{kl})^2]. 
\end{equation}
With this new parametrization, the reactor antineutrino flux will be 
given by 
\begin{equation}
\lambda_{l}(E_\nu) = \exp\big[ \sum_{k=1}^6 c_{kl}p_k^{l}(E_\nu)\big], 
\end{equation}
with $p_k^{l}(E_\nu)$ the polynomial 
\begin{equation}
p_k^{l}(E_\nu) = \sum_{k'=1}^6\mathcal{O}_{kk'}^lE_\nu^{k'-1}.
\end{equation}
After these computations, the covariance matrix will be given as 
\begin{equation}
\sigma_{ij}^2 
= \Delta_i^2 \delta_{ij} + \sum_l\delta N_i^l \delta N_j^l 
\end{equation}
with $\Delta_i^2$ the statistical uncertainty for the $i^{th}$ bin either 
for the CONNIE or CONUS future measurement, and $\delta
N_i^l$ the contribution to the systematic error of the $i^{th}$ bin due
to the $l$ isotope ($^{235}$~U, $^{239}$~Pu, $^{241}$~Pu, $^{238}$~U).
 
With all these computations we can now define the covariance analysis
for the $\chi^2$ function 
\begin{equation}
\chi^2 = \sum_{ij}(N_i^{theo} - N_i^{exp}) \sigma_{ij}^{-2}
(N_j^{theo} - N_j^{exp}) ,
\end{equation}
with 
\begin{equation}
N_i^{theo} = N_i^{235} + N_i^{238} + N_i^{241} + N_i^{239} ,
\end{equation}
the theoretical expected number of events. 

\begin{figure}[t] 
\begin{center}
\includegraphics[width=0.49\textwidth]{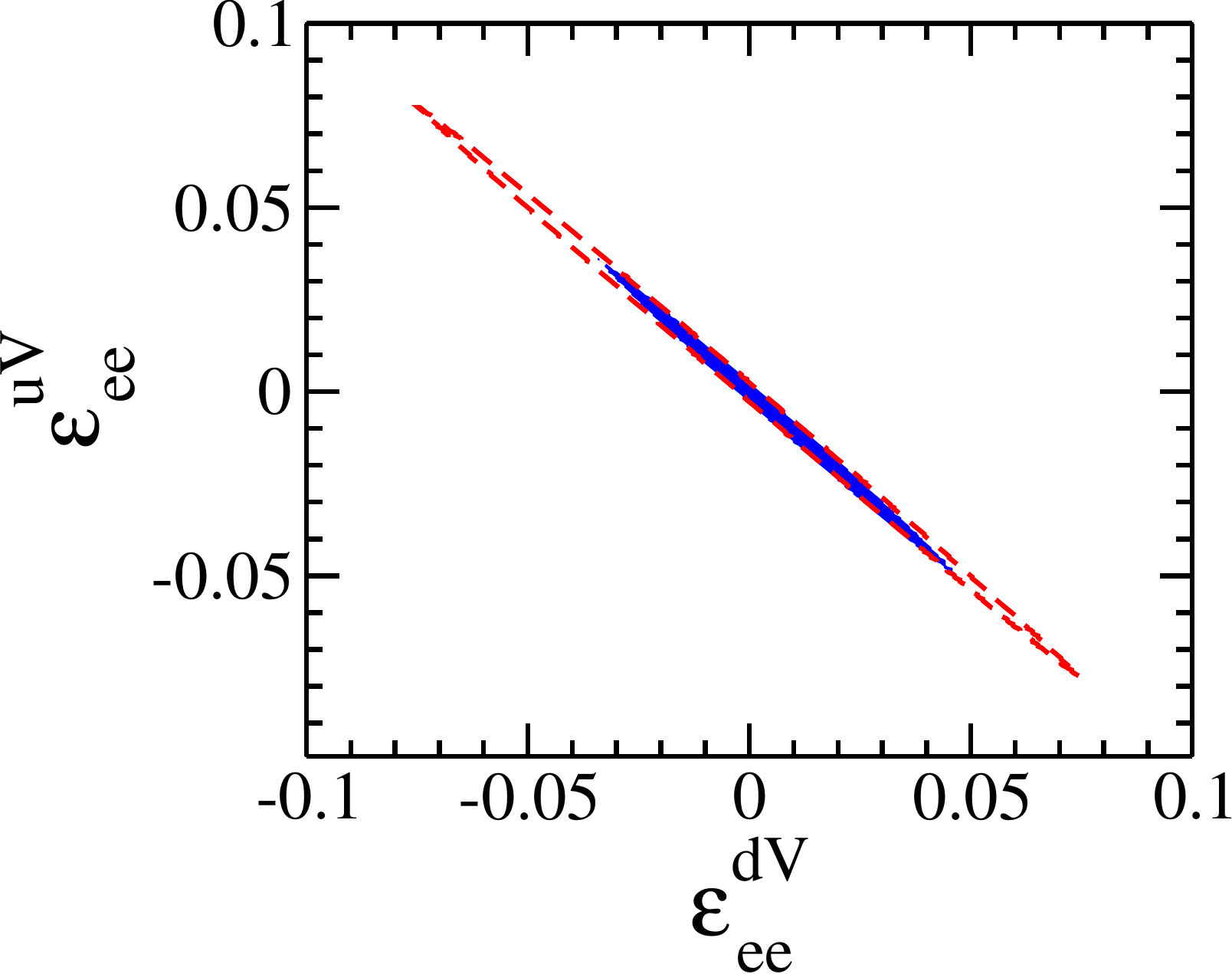}
\end{center}
\caption{Expected exclusion for the non-universal NSI parameters for a
  combined analysis of the reactor neutrino experiments CONNIE and
  CONUS. We include only the statistical error and the correlated
  systematic error from the reactor antineutrino spectrum (solid blue
  area). The region indicates the exclusion at 90 \% C.L. when we
  consider a $100$~\% efficiency. For comparison we also show the
  result of considering a $2$~\% systematic error without any
  correlations (dashed line); the results in both cases are qualitatively
  similar.
\label{fig:combi}}
\end{figure}

The results of this analysis are shown in Fig.~\ref{fig:combi} for the
non-universal parameters. We can see the complementarity between both
reactor measurements and also notice that flux uncertainties related
to the reactor antineutrino flux would not be the most relevant source
of errors.  To have a qualitative idea of the role of correlations, we
have also computed the case in which the only source for systematics
is a $2$~\% error on each experiment, without any
correlation. The result corresponds to the region surrounded by the
dashed line in Fig.~\ref{fig:combi}. As we can see, both regions are
qualitatively similar, having the effect of a narrower constraint when
the correlated errors are considered.

\section{ CEvNS experiments with Spallation Neutron Source}

Besides reactor antineutrinos, we can also make an analysis based on
neutrinos coming from the SNS. That is the case
of the COHERENT Collaboration, which used a CsI detector to make the
first measurement of CEvNS. Here, a relatively high energy
proton beam collides with a mercury target to produce high energy
neutrons. As a result of the interaction, there is also a production
of electron neutrinos, muon neutrinos and muon antineutrinos, each with
a distribution of the form:
\begin{equation}
\frac{\mathrm{dN_{\nu _{\mu }}} }{\mathrm{d}E }=\eta\delta\left ( E-\frac{m_{\pi }^{2}-m_{\mu }^{2}}{2m_{\pi }} \right ).
\label{FluxDelta}
\end{equation}
\newline
\begin{equation}
\frac{\mathrm{dN_{\overline{\nu} _{\mu }}} }{\mathrm{d}E }= \eta \frac{64E^{2}}{m_{\mu }^{3}}\left ( \frac{3}{4}-\frac{E}{m_{\mu }} \right )
\label{FluxMuon}
\end{equation}
\newline
\begin{equation}
\frac{\mathrm{dN_{\nu _{e }}} }{\mathrm{d}E }= \eta \frac{192E^{2}}{m_{\mu }^{3}}\left ( \frac{1}{2}-\frac{E}{m_{\mu }} \right ) ,
\label{FluxElectron}
\end{equation}
\newline
where $\eta = rN_{POT}/4\pi L^{2}$ a normalization factor with $r
=0.08$ being the number of neutrinos per flavor; $N_{POT} = 1.76
\times 10^{23}$, the number of protons on target, and $L$ the distance
from the source to the detector. The total neutrino flux will be given
by the sum of the three contributions. In contrast to neutrinos coming
from nuclear reactors, these neutrinos have a maximum energy of
$\sim$52 MeV and, in consequence, the form factors are no longer
constant and they play an important role in the computation of the
cross section. This means that the experimental data coming from this
measurement are sensitive to nuclear information (see below) as
discussed in Ref.~\cite{Cadeddu:2017etk}. In this section we will
simultaneously study the nuclear neutron distribution of iodine
together with the possibility of having NSI interactions in the
process. First we will make the analysis using the currently available
data from the CsI detector and then we will give a future perspective
for the NaI proposal \cite{Akimov:2018ghi}.
\newline\newline 
 Once the neutrino flux has been specified,
  we can compute the expected number of events,  
which for this experiment is given by:
\begin{equation}
N^{th} = N_{D}\int_{T}A(T)dT\int_{E_{min}}^{52.8 MeV}dE\sum_{a} \frac{\mathrm{d}N_{a}}{\mathrm{d}E} \frac{\mathrm{d} \sigma_{a} }{\mathrm{d} T},
\label{SNS_Events}
\end{equation}
\newline
with $N_{D}$ the number of targets within the detector and $A(T)$ an
acceptance function. The nuclear recoil energy $T$ has a maximum value
which is well approximated by $T_{\rm max}(E_{\nu})\simeq
2E_{\nu}^{2}/M$, while the minimum value is specified by the detector
properties.  The sum runs over the three different neutrino fluxes,
each multiplied by the corresponding cross section, which can be
either the standard or the nonstandard one depending on the NSI
parameters under study.  Also, for all our SNS calculations we have
set the same time window as the one reported by the COHERENT
Collaboration in~\cite{Akimov:2017ade}.
\newline\newline As mentioned before, this neutrino counting
experiment is sensitive to NSI parameters and to the neutron
nuclear distribution. The former explicitly through the cross section
in Eq.~(\ref{CrossN}) and the latter through the form factors in the same
equation. It has been shown that the effect of the form factor in the
theoretical number of events is model independent
\cite{Cadeddu:2017etk}, and thus we will consider both proton and
neutron distributions as given by a symmetrized Fermi
one, that is~\cite{Piekarewicz:2016vbn}:
\begin{equation}
F_{A}^{SF}\left ( q^{2} \right )=\frac{3}{qc_{A}\left [ \left ( qc_{A} \right )^{2}+\left ( \pi qa \right )^{2} \right ]}\left [ \frac{\pi qa}{\textrm{sinh}\left ( \pi qa \right )} \right ]\left [ \frac{\pi qa\, \textrm{sin}\left ( qc_{A} \right )}{\textrm{tanh}\left ( \pi qa \right )}-qc_{A}\, \textrm{cos}\left ( qc_{A} \right ) \right ],
\label{FF}
\end{equation}
\newline
with $A=Z, N$ indicating the form factor for protons and neutrons, respectively; $q^{2} = 2MT$ and, 
in both cases, $a = 0.5233$~fm is a parameter related to the surface thickness.
 On the other hand, the parameter $c_{A}$ contains information about the corresponding rms radius $R_{A}$ through the relation,
\begin{equation}
R_{A}^{2} = \frac{3}{5}c_{A}^{2} + \frac{7}{5}(\pi a)^{2}.
\label{Rc}
\end{equation}    
\newline
 We can use this dependence to get constraints for the neutron rms radius of heavy isotopes on the target material.
This can be done by minimizing the squared function:
\newline
\begin{equation}
\chi ^{2} = \left (\frac{N^{exp} - (1+\alpha)N^{th}(X) - (1+\beta)N^{bg}}{\sigma}  \right )^{2} + \left ( \frac{\alpha}{\sigma_{\alpha}} \right )^{2} + \left ( \frac{\beta}{\sigma_{\beta}} \right )^{2},
\label{chi_sq_SNS}
\end{equation}
\newline
 Here $N^{exp}$ represents the
 measured number of events, $N^{th}(X)$ corresponds to the
predicted one as a function of a set of
parameters $X$, $N^{bg}$ is the number of background events coming, for instance, 
from neutrino induced neutrons~\cite{Kolbe:2000np} and prompt neutrons~\cite{Akimov:2017ade},
and $\alpha, \beta$ are parameters which quantify systematic and background errors with their 
corresponding uncertainty.
\newline\newline

\begin{figure}[t] 
\begin{center}
\includegraphics[width=0.49\textwidth]{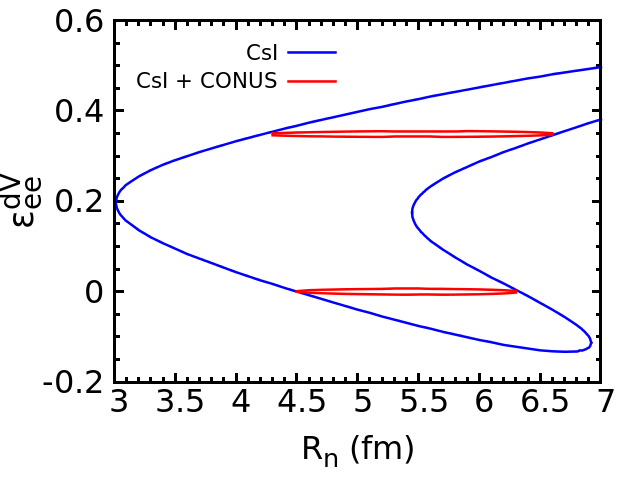}
\includegraphics[width=0.49\textwidth]{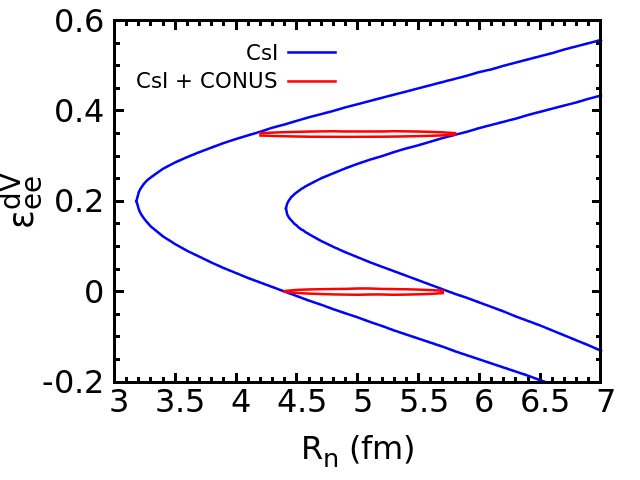}
\end{center}
\caption{\label{fig:NaISNS2} Left panel: Allowed region from the CsI
  COHERENT data for the NSI parameter
  $\epsilon^{dV}_{ee}$ vs the mean neutron radius $R_n$ at
  $1\sigma$. The original quenching factor reported by the
  Collaboration was considered in this case. The
  restriction from a combined result with a future measure from the
  CONUS reactor experiment is illustrated with a red line. Right
  panel: same analysis as in the left panel, but considering now the
  recent reported new quenching factor (see text for details). In both panels we have
  considered that reactors experiments will have an $8$~\% systematic
  error.} 
\end{figure}

Based on the previous discussion we present our results. First, we
will present our obtained constraints from the published CsI COHERENT
data~\cite{Akimov:2017ade}. To this end, by following the approach in Ref.  \cite{Cadeddu:2017etk}, we will consider Cs and I nuclei as indistinguishable, with their
corresponding proton distributions as is well known.
Besides, the acceptance function in Eq. (\ref{SNS_Events}) will be given by 

\begin{equation}
A(x) = \frac{a}{1 + \textup{exp}(-k(x-x_{0}))} \Theta (x-5),
\label{Acceptance}
\end{equation}
\newline
where $\Theta$ is a Heaviside modified function and $a$, $k$, and
$x_{0}$ are fixed parameters \cite{Akimov:2018vzs}.  The argument $x$
(number of photoelectrons) in Eq. (\ref{Acceptance}) satisfies
$x~\propto~Q_{f}(T_{A})T_{A}$, with $Q_{f}$ the quenching factor,
defined as the nuclear recoil energy fraction, compared to an electron
recoil. Initially, this quenching factor (QF) was considered by the
COHERENT Collaboration as constant \cite{Akimov:2017ade}, but recent
studies have proposed a behavior as a function of the nuclear recoil
energy as shown in Ref. \cite{Collar:2019ihs}. For our calculations we
will consider both pictures and see how the results are affected by
the choice of QF.
\newline\newline
    Besides being sensitive to $R_{n}$, notice that
 the cross section in (\ref{CrossN}) also depends on NSI parameters. This means that any experimentally measured deviation from the
SM number of events in Eq. (\ref{SNS_Events}) can be attributed to a lack of knowledge of the nuclear distribution
of the Cs and I nuclei or to new physics. Motivated by this last statement, 
and as a first approach, we will consider one of the NSI parameters, $\epsilon_{ee}^{dV}$,  
as nonzero and we will constrain it together with 
$R_{n}$. That is, we will minimize
Eq. (\ref{chi_sq_SNS}) with $X = \{ \epsilon_{ee}^{dV}, R_{n}\}$.
\newline\newline The analysis in this case was carried out by energy
bins as in \cite{Cadeddu:2017etk}, with experimental data and
background events, and their corresponding uncertainties, taken from
\cite{Akimov:2017ade}.  The results are shown at 1$\sigma$ (region
between blue lines) in Fig. \ref{fig:NaISNS2}, where a comparison is
done for different QF. The left panel of Fig.~\ref{fig:NaISNS2} 
shows the results with the QF used by the COHERENT Collaboration,
which implies $\sigma_{\alpha} = 0.28$, whereas in the right panel 
we used the QF proposed in \cite{Collar:2019ihs}, with
$\sigma_{\alpha} = 0.135$. In both cases we took $\sigma_{\beta} =
0.25$.  Notice that the effect of taking a different QF is observed as
a displacement in the allowed region. 
\newline\newline In addition, motivated by the discussion in
Sect~\ref{sec:1}, we can combine the results of nuclear reactor and
SNS experiments to get better constraints to both parameters.  To this
end we follow a simple approach for which the total $\chi^{2}$
function will be given by the sum of the individual ones, as usual for
independent observables.  Due to the low energy regime, the reactor
measurements will be only sensitive to the NSI interactions.  The
results are also shown in the same left and right panels of
Fig.~\ref{fig:NaISNS2}, respectively, where we present the expected region of the
combined analysis of the COHERENT data with the expected results from
the CONUS experiment. In the first panel, the allowed region is
consistent with Ref. \cite{Cadeddu:2017etk}, which corresponds to the
standard case without NSI.
\newline

 The complete plan of the COHERENT Collaboration includes a NaI
 detector to measure CEvNS. We will show what would be the expected
 impact on the previous results by using this kind of detector. To this end,
we follow a similar approach as for the CsI detector, but as we are talking about a
 future experiment, we need to make some assumptions.  We will
 take the material detector mass, distance from the neutrino source,
 and threshold nuclear recoil energy as those given in
 Ref. \cite{Akimov:2018ghi}, and regarding the acceptance function we
 will consider the ideal case on which it is equal to one for all $T$,
 this is justified by the fact that the characterization of the NaI
 material has not been reported yet.  Another important assumption is
 that we will take the experimental number of events in
 Eq.~(\ref{SNS_Events}) to be the SM prediction considering a neutron
 rms radius of 5.5 fm, which corresponds to the minimum in the squared
 function when there are no NSI parameters involved and when a QF as
 that used by the COHERENT Collaboration is taken. Finally, we
 considered an optimistic case on which $\sigma_{\alpha} = 0.05$ with
 $N^{bg} = 10\%$ of the SM prediction and $\sigma_{\beta} = 0.10$.
\newline\newline
  \begin{figure}[t] 
\begin{center}
\includegraphics[width=0.49\textwidth]{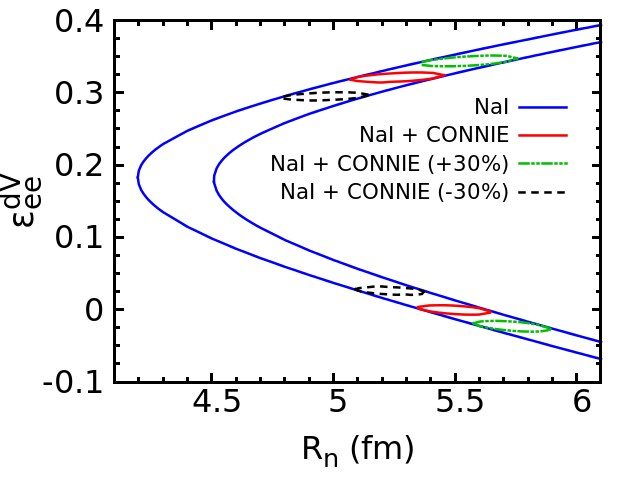}
\end{center}
\caption{\label{fig:NaISNS3} Expected futuristic constraints from a NaI detector in
  combination with a CONNIE futuristic result at $1\sigma$. In this
  case, we have considered different cases for the total measured
  events at CONNIE that illustrate how the combination of the two
  experiments can discriminate the values of both standard and
  nonstandard parameters. We have
  considered that reactors experiments will have an $8$~\% systematic
  error.} 
\end{figure}
The results are shown at a 1$\sigma$ level as the region between
blue lines in Fig.~\ref{fig:NaISNS3}. Following the previous analysis, we also show the expected
region, combined with the CONNIE experiment
analysis from Sec.~\ref{sec:3}. In addition, we show the expected results
 assuming that the CONNIE experiment, (the one with
  higher systematic uncertainties) reports a $\pm$30\%
deviation from the SM prediction. This
  assumption is only illustrative, so we can see that a considerable
      deviation from the SM prediction results in
  a displacement of the central value related to both the nuclear rms
  radius and the NSI parameter under study, but with the width of the
  errors unchanged. Notice that if CONNIE measures
an excess above the SM prediction, the average radius should have a larger
value, while the NSI signal will have negative values.  In
contrast, when the measured number of events for CONNIE is lower, the
neutron radius takes lower values and the NSI parameter prefers larger ones.
This is consistent with the form factor dependence on the
neutron radius and with the cross section linear and quadratic dependence on
 the non-universal NSI parameter $\epsilon^{dV}_{ee}$.

\section{Conclusions}
The increasing activity for a first detection of CEvNS with reactor
antineutrinos and for an improved measurement of SNS
CEvNS~\cite{Akimov:2018ghi,Baxter:2019mcx}, opens the door to future precise
neutrino physics and to  physics beyond the Standard
Model constraints.

In order to have reliable constraints on physics beyond the Standard
Model, we should have under control all possible standard
contributions for a variation in the number of events, including the
freedom in the nuclear physics parameters that appear in the neutron
form factor. We have illustrated how current measurement of CEvNS can
be ascribed either  to a correction in the neutron mean radius or to a
variation in the NSI parameters. In the near
future, reactor and SNS CEvNS measurements could be complementary and
will help to improve the knowledge of nuclear and Standard Model parameters, 
constraining at the same time  new physics, such as NSI parameters.

We have computed the expected sensitivity to NSI for different CEvNS
proposals in reactor neutrinos and showed how, in combination with SNS
experiments, they can contribute to have robust measurement of the mean
neutron radius while improving NSI constraints at the same time.

\acknowledgments { This work was supported by CONACYT-Mexico grant
  A1-S-23238, SNI (Sistema Nacional de Investigadores), and PAPIIT
  project IN115319. A. Parada was supported by Universidad Santiago de
  Cali (USC) under grant 935-621118-3. B. C. Canas was supported by
  Universidad de Pamplona under grant 400-156.012-027(GA313-BP-2019)}


\begin{thebibliography}{88}
\expandafter\ifx\csname natexlab\endcsname\relax\def\natexlab#1{#1}\fi
\expandafter\ifx\csname bibnamefont\endcsname\relax
  \def\bibnamefont#1{#1}\fi
\expandafter\ifx\csname bibfnamefont\endcsname\relax
  \def\bibfnamefont#1{#1}\fi
\expandafter\ifx\csname citenamefont\endcsname\relax
  \def\citenamefont#1{#1}\fi
\expandafter\ifx\csname url\endcsname\relax
  \def\url#1{\texttt{#1}}\fi
\expandafter\ifx\csname urlprefix\endcsname\relax\def\urlprefix{URL }\fi
\providecommand{\bibinfo}[2]{#2}
\providecommand{\eprint}[2][]{\url{#2}}

\bibitem[{\citenamefont{Freedman}(1974)}]{Freedman:1973yd}
\bibinfo{author}{\bibfnamefont{D.~Z.} \bibnamefont{Freedman}},
  \bibinfo{journal}{Phys. Rev.} \textbf{\bibinfo{volume}{D9}},
  \bibinfo{pages}{1389} (\bibinfo{year}{1974}).

\bibitem[{\citenamefont{Akimov et~al.}(2017)}]{Akimov:2017ade}
\bibinfo{author}{\bibfnamefont{D.}~\bibnamefont{Akimov}} \bibnamefont{et~al.}
  (\bibinfo{collaboration}{COHERENT}), \bibinfo{journal}{Science}
  \textbf{\bibinfo{volume}{357}}, \bibinfo{pages}{1123} (\bibinfo{year}{2017}),
  \eprint{1708.01294}.

\bibitem[{\citenamefont{Papoulias}(2019)}]{Papoulias:2019txv}
\bibinfo{author}{\bibfnamefont{D.~K.} \bibnamefont{Papoulias}}
  (\bibinfo{year}{2019}), \eprint{1907.11644}.

\bibitem[{\citenamefont{Khan and Rodejohann}(2019)}]{Khan:2019cvi}
\bibinfo{author}{\bibfnamefont{A.~N.} \bibnamefont{Khan}} \bibnamefont{and}
  \bibinfo{author}{\bibfnamefont{W.}~\bibnamefont{Rodejohann}}
  \bibinfo{journal}{Phys. Rev.} \textbf{\bibinfo{volume}{D100}},
  \bibinfo{pages}{113003} (\bibinfo{year}{2019}).
  \eprint{1907.12444}.

\bibitem[{\citenamefont{Coloma et~al.}(2019)\citenamefont{Coloma, Esteban,
  Gonzalez-Garcia, and Maltoni}}]{Coloma:2019mbs}
\bibinfo{author}{\bibfnamefont{P.}~\bibnamefont{Coloma}},
  \bibinfo{author}{\bibfnamefont{I.}~\bibnamefont{Esteban}},
  \bibinfo{author}{\bibfnamefont{M.~C.} \bibnamefont{Gonzalez-Garcia}},
  \bibnamefont{and} \bibinfo{author}{\bibfnamefont{M.}~\bibnamefont{Maltoni}}
  (\bibinfo{year}{2019}), \eprint{1911.09109}.

\bibitem[{\citenamefont{Papoulias
  et~al.}(2019{\natexlab{a}})\citenamefont{Papoulias, Kosmas, and
  Kuno}}]{Papoulias:2019xaw}
\bibinfo{author}{\bibfnamefont{D.~K.} \bibnamefont{Papoulias}},
  \bibinfo{author}{\bibfnamefont{T.~S.} \bibnamefont{Kosmas}},
  \bibnamefont{and} \bibinfo{author}{\bibfnamefont{Y.}~\bibnamefont{Kuno}}
  \bibinfo{journal}{Front. Phys.} \textbf{\bibinfo{volume}{7}},
  \bibinfo{pages}{191} (\bibinfo{year}{2019}),
  \eprint{1911.00916}

\bibitem[{\citenamefont{Aristizabal~Sierra
  et~al.}(2019{\natexlab{a}})\citenamefont{Aristizabal~Sierra, Dutta, Liao, and
  Strigari}}]{AristizabalSierra:2019ykk}
\bibinfo{author}{\bibfnamefont{D.}~\bibnamefont{Aristizabal~Sierra}},
  \bibinfo{author}{\bibfnamefont{B.}~\bibnamefont{Dutta}},
  \bibinfo{author}{\bibfnamefont{S.}~\bibnamefont{Liao}}, \bibnamefont{and}
  \bibinfo{author}{\bibfnamefont{L.~E.} \bibnamefont{Strigari}}
  \bibinfo{journal}{J. High Energy Phys.} \textbf{\bibinfo{volume}{19:12}},
  \bibinfo{pages}{124} (\bibinfo{year}{2019}),
  \eprint{1910.12437}.

\bibitem[{\citenamefont{Miranda
  et~al.}(2019{\natexlab{a}})\citenamefont{Miranda, Sanchez~Garcia, and
  Sanders}}]{Miranda:2019skf}
\bibinfo{author}{\bibfnamefont{O.~G.} \bibnamefont{Miranda}},
  \bibinfo{author}{\bibfnamefont{G.}~\bibnamefont{Sanchez~Garcia}},
  \bibnamefont{and} \bibinfo{author}{\bibfnamefont{O.}~\bibnamefont{Sanders}},
  \bibinfo{journal}{Adv. High Energy Phys.} \textbf{\bibinfo{volume}{2019}},
  \bibinfo{pages}{3902819} (\bibinfo{year}{2019}{\natexlab{a}}),
  \eprint{1902.09036}.

\bibitem[{\citenamefont{Altmannshofer et~al.}(2019)\citenamefont{Altmannshofer,
  Tammaro, and Zupan}}]{Altmannshofer:2018xyo}
\bibinfo{author}{\bibfnamefont{W.}~\bibnamefont{Altmannshofer}},
  \bibinfo{author}{\bibfnamefont{M.}~\bibnamefont{Tammaro}}, \bibnamefont{and}
  \bibinfo{author}{\bibfnamefont{J.}~\bibnamefont{Zupan}},
  \bibinfo{journal}{JHEP} \textbf{\bibinfo{volume}{09}}, \bibinfo{pages}{083}
  (\bibinfo{year}{2019}), \eprint{1812.02778}.

\bibitem[{\citenamefont{Cadeddu
  et~al.}(2018{\natexlab{a}})\citenamefont{Cadeddu, Giunti, Kouzakov, Li,
  Studenikin, and Zhang}}]{Cadeddu:2018dux}
\bibinfo{author}{\bibfnamefont{M.}~\bibnamefont{Cadeddu}},
  \bibinfo{author}{\bibfnamefont{C.}~\bibnamefont{Giunti}},
  \bibinfo{author}{\bibfnamefont{K.~A.} \bibnamefont{Kouzakov}},
  \bibinfo{author}{\bibfnamefont{Y.~F.} \bibnamefont{Li}},
  \bibinfo{author}{\bibfnamefont{A.~I.} \bibnamefont{Studenikin}},
  \bibnamefont{and} \bibinfo{author}{\bibfnamefont{Y.~Y.} \bibnamefont{Zhang}},
  \bibinfo{journal}{Phys. Rev.} \textbf{\bibinfo{volume}{D98}},
  \bibinfo{pages}{113010} (\bibinfo{year}{2018}{\natexlab{a}}),
  \eprint{1810.05606}.

\bibitem[{\citenamefont{Ca\~nas
  et~al.}(2018{\natexlab{a}})\citenamefont{Ca\~nas, Garc\'es, Miranda, and
  Parada}}]{Canas:2018rng}
\bibinfo{author}{\bibfnamefont{B.~C.} \bibnamefont{Ca\~nas}},
  \bibinfo{author}{\bibfnamefont{E.~A.} \bibnamefont{Garc\'es}},
  \bibinfo{author}{\bibfnamefont{O.~G.} \bibnamefont{Miranda}},
  \bibnamefont{and} \bibinfo{author}{\bibfnamefont{A.}~\bibnamefont{Parada}},
  \bibinfo{journal}{Phys. Lett.} \textbf{\bibinfo{volume}{B784}},
  \bibinfo{pages}{159} (\bibinfo{year}{2018}{\natexlab{a}}),
  \eprint{1806.01310}.

\bibitem[{\citenamefont{Billard et~al.}(2018)\citenamefont{Billard, Johnston,
  and Kavanagh}}]{Billard:2018jnl}
\bibinfo{author}{\bibfnamefont{J.}~\bibnamefont{Billard}},
  \bibinfo{author}{\bibfnamefont{J.}~\bibnamefont{Johnston}}, \bibnamefont{and}
  \bibinfo{author}{\bibfnamefont{B.~J.} \bibnamefont{Kavanagh}},
  \bibinfo{journal}{JCAP} \textbf{\bibinfo{volume}{1811}}, \bibinfo{pages}{016}
  (\bibinfo{year}{2018}), \eprint{1805.01798}.

\bibitem[{\citenamefont{Denton et~al.}(2018)\citenamefont{Denton, Farzan, and
  Shoemaker}}]{Denton:2018xmq}
\bibinfo{author}{\bibfnamefont{P.~B.} \bibnamefont{Denton}},
  \bibinfo{author}{\bibfnamefont{Y.}~\bibnamefont{Farzan}}, \bibnamefont{and}
  \bibinfo{author}{\bibfnamefont{I.~M.} \bibnamefont{Shoemaker}},
  \bibinfo{journal}{JHEP} \textbf{\bibinfo{volume}{07}}, \bibinfo{pages}{037}
  (\bibinfo{year}{2018}), \eprint{1804.03660}.

\bibitem[{\citenamefont{Farzan et~al.}(2018)\citenamefont{Farzan, Lindner,
  Rodejohann, and Xu}}]{Farzan:2018gtr}
\bibinfo{author}{\bibfnamefont{Y.}~\bibnamefont{Farzan}},
  \bibinfo{author}{\bibfnamefont{M.}~\bibnamefont{Lindner}},
  \bibinfo{author}{\bibfnamefont{W.}~\bibnamefont{Rodejohann}},
  \bibnamefont{and} \bibinfo{author}{\bibfnamefont{X.-J.} \bibnamefont{Xu}},
  \bibinfo{journal}{JHEP} \textbf{\bibinfo{volume}{05}}, \bibinfo{pages}{066}
  (\bibinfo{year}{2018}), \eprint{1802.05171}.

\bibitem[{\citenamefont{Papoulias and Kosmas}(2018)}]{Kosmas:2017tsq}
\bibinfo{author}{\bibfnamefont{D.~K.} \bibnamefont{Papoulias}}
  \bibnamefont{and} \bibinfo{author}{\bibfnamefont{T.~S.}
  \bibnamefont{Kosmas}}, \bibinfo{journal}{Phys. Rev.}
  \textbf{\bibinfo{volume}{D97}}, \bibinfo{pages}{033003}
  (\bibinfo{year}{2018}), \eprint{1711.09773}.

\bibitem[{\citenamefont{Cadeddu
  et~al.}(2018{\natexlab{b}})\citenamefont{Cadeddu, Giunti, Li, and
  Zhang}}]{Cadeddu:2017etk}
\bibinfo{author}{\bibfnamefont{M.}~\bibnamefont{Cadeddu}},
  \bibinfo{author}{\bibfnamefont{C.}~\bibnamefont{Giunti}},
  \bibinfo{author}{\bibfnamefont{Y.~F.} \bibnamefont{Li}}, \bibnamefont{and}
  \bibinfo{author}{\bibfnamefont{Y.~Y.} \bibnamefont{Zhang}},
  \bibinfo{journal}{Phys. Rev. Lett.} \textbf{\bibinfo{volume}{120}},
  \bibinfo{pages}{072501} (\bibinfo{year}{2018}{\natexlab{b}}),
  \eprint{1710.02730}.

\bibitem[{\citenamefont{Ca\~nas
  et~al.}(2018{\natexlab{b}})\citenamefont{Ca\~nas, Garc\'es, Miranda, and
  Parada}}]{Canas:2017umu}
\bibinfo{author}{\bibfnamefont{B.~C.} \bibnamefont{Ca\~nas}},
  \bibinfo{author}{\bibfnamefont{E.~A.} \bibnamefont{Garc\'es}},
  \bibinfo{author}{\bibfnamefont{O.~G.} \bibnamefont{Miranda}},
  \bibnamefont{and} \bibinfo{author}{\bibfnamefont{A.}~\bibnamefont{Parada}},
  \bibinfo{journal}{Phys. Lett.} \textbf{\bibinfo{volume}{B776}},
  \bibinfo{pages}{451} (\bibinfo{year}{2018}{\natexlab{b}}),
  \eprint{1708.09518}.

\bibitem[{\citenamefont{Ge and Shoemaker}(2018)}]{Ge:2017mcq}
\bibinfo{author}{\bibfnamefont{S.-F.} \bibnamefont{Ge}} \bibnamefont{and}
  \bibinfo{author}{\bibfnamefont{I.~M.} \bibnamefont{Shoemaker}},
  \bibinfo{journal}{JHEP} \textbf{\bibinfo{volume}{11}}, \bibinfo{pages}{066}
  (\bibinfo{year}{2018}), \eprint{1710.10889}.

\bibitem[{\citenamefont{Garces et~al.}(2017)\citenamefont{Garces, Cañas,
  Miranda, and Parada}}]{Garces:2017qkm}
\bibinfo{author}{\bibfnamefont{E.}~\bibnamefont{Garces}},
  \bibinfo{author}{\bibfnamefont{B.}~\bibnamefont{Cañas}},
  \bibinfo{author}{\bibfnamefont{O.}~\bibnamefont{Miranda}}, \bibnamefont{and}
  \bibinfo{author}{\bibfnamefont{A.}~\bibnamefont{Parada}},
  \bibinfo{journal}{J. Phys. Conf. Ser.} \textbf{\bibinfo{volume}{934}},
  \bibinfo{pages}{012004} (\bibinfo{year}{2017}).

\bibitem[{\citenamefont{Shoemaker}(2017)}]{Shoemaker:2017lzs}
\bibinfo{author}{\bibfnamefont{I.~M.} \bibnamefont{Shoemaker}},
  \bibinfo{journal}{Phys. Rev.} \textbf{\bibinfo{volume}{D95}},
  \bibinfo{pages}{115028} (\bibinfo{year}{2017}), \eprint{1703.05774}.

\bibitem[{\citenamefont{Kosmas et~al.}(2017)\citenamefont{Kosmas, Papoulias,
  Tortola, and Valle}}]{Kosmas:2017zbh}
\bibinfo{author}{\bibfnamefont{T.~S.} \bibnamefont{Kosmas}},
  \bibinfo{author}{\bibfnamefont{D.~K.} \bibnamefont{Papoulias}},
  \bibinfo{author}{\bibfnamefont{M.}~\bibnamefont{Tortola}}, \bibnamefont{and}
  \bibinfo{author}{\bibfnamefont{J.~W.~F.} \bibnamefont{Valle}},
  \bibinfo{journal}{Phys. Rev.} \textbf{\bibinfo{volume}{D96}},
  \bibinfo{pages}{063013} (\bibinfo{year}{2017}), \eprint{1703.00054}.

\bibitem[{\citenamefont{Dutta et~al.}(2016)\citenamefont{Dutta, Gao, Mahapatra,
  Mirabolfathi, Strigari, and Walker}}]{Dutta:2015nlo}
\bibinfo{author}{\bibfnamefont{B.}~\bibnamefont{Dutta}},
  \bibinfo{author}{\bibfnamefont{Y.}~\bibnamefont{Gao}},
  \bibinfo{author}{\bibfnamefont{R.}~\bibnamefont{Mahapatra}},
  \bibinfo{author}{\bibfnamefont{N.}~\bibnamefont{Mirabolfathi}},
  \bibinfo{author}{\bibfnamefont{L.~E.} \bibnamefont{Strigari}},
  \bibnamefont{and} \bibinfo{author}{\bibfnamefont{J.~W.}
  \bibnamefont{Walker}}, \bibinfo{journal}{Phys. Rev.}
  \textbf{\bibinfo{volume}{D94}}, \bibinfo{pages}{093002}
  (\bibinfo{year}{2016}), \eprint{1511.02834}.

\bibitem[{\citenamefont{Kosmas et~al.}(2015{\natexlab{a}})\citenamefont{Kosmas,
  Miranda, Papoulias, Tortola, and Valle}}]{Kosmas:2015vsa}
\bibinfo{author}{\bibfnamefont{T.~S.} \bibnamefont{Kosmas}},
  \bibinfo{author}{\bibfnamefont{O.~G.} \bibnamefont{Miranda}},
  \bibinfo{author}{\bibfnamefont{D.~K.} \bibnamefont{Papoulias}},
  \bibinfo{author}{\bibfnamefont{M.}~\bibnamefont{Tortola}}, \bibnamefont{and}
  \bibinfo{author}{\bibfnamefont{J.~W.~F.} \bibnamefont{Valle}},
  \bibinfo{journal}{Phys. Lett.} \textbf{\bibinfo{volume}{B750}},
  \bibinfo{pages}{459} (\bibinfo{year}{2015}{\natexlab{a}}),
  \eprint{1506.08377}.

\bibitem[{\citenamefont{Kosmas et~al.}(2015{\natexlab{b}})\citenamefont{Kosmas,
  Miranda, Papoulias, Tortola, and Valle}}]{Kosmas:2015sqa}
\bibinfo{author}{\bibfnamefont{T.~S.} \bibnamefont{Kosmas}},
  \bibinfo{author}{\bibfnamefont{O.~G.} \bibnamefont{Miranda}},
  \bibinfo{author}{\bibfnamefont{D.~K.} \bibnamefont{Papoulias}},
  \bibinfo{author}{\bibfnamefont{M.}~\bibnamefont{Tortola}}, \bibnamefont{and}
  \bibinfo{author}{\bibfnamefont{J.~W.~F.} \bibnamefont{Valle}},
  \bibinfo{journal}{Phys. Rev.} \textbf{\bibinfo{volume}{D92}},
  \bibinfo{pages}{013011} (\bibinfo{year}{2015}{\natexlab{b}}),
  \eprint{1505.03202}.

\bibitem[{\citenamefont{Barranco et~al.}(2012)\citenamefont{Barranco, Bolanos,
  Garces, Miranda, and Rashba}}]{Barranco:2011wx}
\bibinfo{author}{\bibfnamefont{J.}~\bibnamefont{Barranco}},
  \bibinfo{author}{\bibfnamefont{A.}~\bibnamefont{Bolanos}},
  \bibinfo{author}{\bibfnamefont{E.~A.} \bibnamefont{Garces}},
  \bibinfo{author}{\bibfnamefont{O.~G.} \bibnamefont{Miranda}},
  \bibnamefont{and} \bibinfo{author}{\bibfnamefont{T.~I.}
  \bibnamefont{Rashba}}, \bibinfo{journal}{Int. J. Mod. Phys.}
  \textbf{\bibinfo{volume}{A27}}, \bibinfo{pages}{1250147}
  (\bibinfo{year}{2012}), \eprint{1108.1220}.

\bibitem[{\citenamefont{Barranco et~al.}(2007)\citenamefont{Barranco, Miranda,
  and Rashba}}]{Barranco:2007tz}
\bibinfo{author}{\bibfnamefont{J.}~\bibnamefont{Barranco}},
  \bibinfo{author}{\bibfnamefont{O.~G.} \bibnamefont{Miranda}},
  \bibnamefont{and} \bibinfo{author}{\bibfnamefont{T.~I.}
  \bibnamefont{Rashba}}, \bibinfo{journal}{Phys. Rev.}
  \textbf{\bibinfo{volume}{D76}}, \bibinfo{pages}{073008}
  (\bibinfo{year}{2007}), \eprint{hep-ph/0702175}.

\bibitem[{\citenamefont{Billard et~al.}(2015)\citenamefont{Billard, Strigari,
  and Figueroa-Feliciano}}]{Billard:2014yka}
\bibinfo{author}{\bibfnamefont{J.}~\bibnamefont{Billard}},
  \bibinfo{author}{\bibfnamefont{L.~E.} \bibnamefont{Strigari}},
  \bibnamefont{and}
  \bibinfo{author}{\bibfnamefont{E.}~\bibnamefont{Figueroa-Feliciano}},
  \bibinfo{journal}{Phys. Rev.} \textbf{\bibinfo{volume}{D91}},
  \bibinfo{pages}{095023} (\bibinfo{year}{2015}), \eprint{1409.0050}.

\bibitem[{\citenamefont{Boehm et~al.}(2019)\citenamefont{Boehm, Cerdeno,
  Machado, Olivares-Del~Campo, and Reid}}]{Boehm:2018sux}
\bibinfo{author}{\bibfnamefont{C.}~\bibnamefont{Boehm}},
  \bibinfo{author}{\bibfnamefont{D.~G.} \bibnamefont{Cerdeno}},
  \bibinfo{author}{\bibfnamefont{P.~A.~N.} \bibnamefont{Machado}},
  \bibinfo{author}{\bibfnamefont{A.}~\bibnamefont{Olivares-Del~Campo}},
  \bibnamefont{and} \bibinfo{author}{\bibfnamefont{E.}~\bibnamefont{Reid}},
  \bibinfo{journal}{JCAP} \textbf{\bibinfo{volume}{1901}}, \bibinfo{pages}{043}
  (\bibinfo{year}{2019}), \eprint{1809.06385}.

\bibitem[{\citenamefont{Papoulias et~al.}(2018)\citenamefont{Papoulias, Sahu,
  Kosmas, Kota, and Nayak}}]{Papoulias:2018uzy}
\bibinfo{author}{\bibfnamefont{D.~K.} \bibnamefont{Papoulias}},
  \bibinfo{author}{\bibfnamefont{R.}~\bibnamefont{Sahu}},
  \bibinfo{author}{\bibfnamefont{T.~S.} \bibnamefont{Kosmas}},
  \bibinfo{author}{\bibfnamefont{V.~K.~B.} \bibnamefont{Kota}},
  \bibnamefont{and} \bibinfo{author}{\bibfnamefont{B.}~\bibnamefont{Nayak}},
  \bibinfo{journal}{Adv. High Energy Phys.} \textbf{\bibinfo{volume}{2018}},
  \bibinfo{pages}{6031362} (\bibinfo{year}{2018}), \eprint{1804.11319}.

\bibitem[{\citenamefont{Aristizabal~Sierra
  et~al.}(2018{\natexlab{a}})\citenamefont{Aristizabal~Sierra, Rojas, and
  Tytgat}}]{AristizabalSierra:2017joc}
\bibinfo{author}{\bibfnamefont{D.}~\bibnamefont{Aristizabal~Sierra}},
  \bibinfo{author}{\bibfnamefont{N.}~\bibnamefont{Rojas}}, \bibnamefont{and}
  \bibinfo{author}{\bibfnamefont{M.~H.~G.} \bibnamefont{Tytgat}},
  \bibinfo{journal}{JHEP} \textbf{\bibinfo{volume}{03}}, \bibinfo{pages}{197}
  (\bibinfo{year}{2018}{\natexlab{a}}), \eprint{1712.09667}.

\bibitem[{\citenamefont{Wong et~al.}(2006)\citenamefont{Wong, Li, Li, Yue, and
  Zhou}}]{Wong:2005vg}
\bibinfo{author}{\bibfnamefont{H.~T.} \bibnamefont{Wong}},
  \bibinfo{author}{\bibfnamefont{H.-B.} \bibnamefont{Li}},
  \bibinfo{author}{\bibfnamefont{J.}~\bibnamefont{Li}},
  \bibinfo{author}{\bibfnamefont{Q.}~\bibnamefont{Yue}}, \bibnamefont{and}
  \bibinfo{author}{\bibfnamefont{Z.-Y.} \bibnamefont{Zhou}},
  \bibinfo{journal}{J. Phys. Conf. Ser.} \textbf{\bibinfo{volume}{39}},
  \bibinfo{pages}{266} (\bibinfo{year}{2006}), \bibinfo{note}{[,344(2005)]},
  \eprint{hep-ex/0511001}.

\bibitem[{\citenamefont{Lindner et~al.}(2017)\citenamefont{Lindner, Rodejohann,
  and Xu}}]{Lindner:2016wff}
\bibinfo{author}{\bibfnamefont{M.}~\bibnamefont{Lindner}},
  \bibinfo{author}{\bibfnamefont{W.}~\bibnamefont{Rodejohann}},
  \bibnamefont{and} \bibinfo{author}{\bibfnamefont{X.-J.} \bibnamefont{Xu}},
  \bibinfo{journal}{JHEP} \textbf{\bibinfo{volume}{03}}, \bibinfo{pages}{097}
  (\bibinfo{year}{2017}), \eprint{1612.04150}.

\bibitem[{\citenamefont{Rothe}(2018)}]{Rothe:2018ulo}
\bibinfo{author}{\bibfnamefont{J.}~\bibnamefont{Rothe}}
  (\bibinfo{collaboration}{NU-CLEUS}), \bibinfo{journal}{PoS}
  \textbf{\bibinfo{volume}{NOW2018}}, \bibinfo{pages}{092}
  (\bibinfo{year}{2018}).

\bibitem[{\citenamefont{Aguilar-Arevalo
  et~al.}(2016{\natexlab{a}})}]{Aguilar-Arevalo:2016qen}
\bibinfo{author}{\bibfnamefont{A.}~\bibnamefont{Aguilar-Arevalo}}
  \bibnamefont{et~al.} (\bibinfo{collaboration}{CONNIE}),
  \bibinfo{journal}{JINST} \textbf{\bibinfo{volume}{11}},
  \bibinfo{pages}{P07024} (\bibinfo{year}{2016}{\natexlab{a}}),
  \eprint{1604.01343}.

\bibitem[{\citenamefont{Aguilar-Arevalo
  et~al.}(2016{\natexlab{b}})}]{Aguilar-Arevalo:2016khx}
\bibinfo{author}{\bibfnamefont{A.}~\bibnamefont{Aguilar-Arevalo}}
  \bibnamefont{et~al.} (\bibinfo{collaboration}{CONNIE}), \bibinfo{journal}{J.
  Phys. Conf. Ser.} \textbf{\bibinfo{volume}{761}}, \bibinfo{pages}{012057}
  (\bibinfo{year}{2016}{\natexlab{b}}), \eprint{1608.01565}.

\bibitem[{\citenamefont{Chávez-Estrada and
  Aguilar-Arevalo}(2017)}]{Chavez-Estrada:2017gni}
\bibinfo{author}{\bibfnamefont{M.}~\bibnamefont{Chávez-Estrada}}
  \bibnamefont{and} \bibinfo{author}{\bibfnamefont{A.~A.}
  \bibnamefont{Aguilar-Arevalo}}, \bibinfo{journal}{J. Phys. Conf. Ser.}
  \textbf{\bibinfo{volume}{912}}, \bibinfo{pages}{012031}
  (\bibinfo{year}{2017}).

\bibitem[{\citenamefont{Aguilar-Arevalo
  et~al.}(2019{\natexlab{a}})}]{Aguilar-Arevalo:2019jlr}
\bibinfo{author}{\bibfnamefont{A.}~\bibnamefont{Aguilar-Arevalo}}
  \bibnamefont{et~al.} (\bibinfo{collaboration}{CONNIE}),
  \bibinfo{journal}{Phys. Rev.} \textbf{\bibinfo{volume}{D100}},
  \bibinfo{pages}{092005} (\bibinfo{year}{2019}{\natexlab{a}}),
  \eprint{1906.02200}.

\bibitem[{\citenamefont{Aguilar-Arevalo
  et~al.}(2019{\natexlab{b}})}]{Aguilar-Arevalo:2019zme}
\bibinfo{author}{\bibfnamefont{A.}~\bibnamefont{Aguilar-Arevalo}}
  \bibnamefont{et~al.} (\bibinfo{collaboration}{CONNIE})
  (\bibinfo{year}{2019}{\natexlab{b}}), \eprint{1910.04951}.

\bibitem[{\citenamefont{Agnolet et~al.}(2017)}]{Agnolet:2016zir}
\bibinfo{author}{\bibfnamefont{G.}~\bibnamefont{Agnolet}} \bibnamefont{et~al.}
  (\bibinfo{collaboration}{MINER}), \bibinfo{journal}{Nucl. Instrum. Meth.}
  \textbf{\bibinfo{volume}{A853}}, \bibinfo{pages}{53} (\bibinfo{year}{2017}),
  \eprint{1609.02066}.

\bibitem[{\citenamefont{Akimov et~al.}(2013)}]{Akimov:2012aya}
\bibinfo{author}{\bibfnamefont{D.~{\relax Yu}.} \bibnamefont{Akimov}}
  \bibnamefont{et~al.} (\bibinfo{collaboration}{RED}), \bibinfo{journal}{JINST}
  \textbf{\bibinfo{volume}{8}}, \bibinfo{pages}{P10023} (\bibinfo{year}{2013}),
  \eprint{1212.1938}.

\bibitem[{\citenamefont{Billard et~al.}(2017)}]{Billard:2016giu}
\bibinfo{author}{\bibfnamefont{J.}~\bibnamefont{Billard}} \bibnamefont{et~al.},
  \bibinfo{journal}{J. Phys.} \textbf{\bibinfo{volume}{G44}},
  \bibinfo{pages}{105101} (\bibinfo{year}{2017}), \eprint{1612.09035}.

\bibitem[{\citenamefont{Huang and Chen}(2019)}]{Huang:2019ene}
\bibinfo{author}{\bibfnamefont{X.-R.} \bibnamefont{Huang}} \bibnamefont{and}
  \bibinfo{author}{\bibfnamefont{L.-W.} \bibnamefont{Chen}},
  \bibinfo{journal}{Phys. Rev.} \textbf{\bibinfo{volume}{D100}},
  \bibinfo{pages}{071301} (\bibinfo{year}{2019}), \eprint{1902.07625}.

\bibitem[{\citenamefont{Aristizabal~Sierra
  et~al.}(2019{\natexlab{b}})\citenamefont{Aristizabal~Sierra, Liao, and
  Marfatia}}]{AristizabalSierra:2019zmy}
\bibinfo{author}{\bibfnamefont{D.}~\bibnamefont{Aristizabal~Sierra}},
  \bibinfo{author}{\bibfnamefont{J.}~\bibnamefont{Liao}}, \bibnamefont{and}
  \bibinfo{author}{\bibfnamefont{D.}~\bibnamefont{Marfatia}},
  \bibinfo{journal}{JHEP} \textbf{\bibinfo{volume}{06}}, \bibinfo{pages}{141}
  (\bibinfo{year}{2019}{\natexlab{b}}), \eprint{1902.07398}.

\bibitem[{\citenamefont{Farzan and Tortola}(2018)}]{Farzan:2017xzy}
\bibinfo{author}{\bibfnamefont{Y.}~\bibnamefont{Farzan}} \bibnamefont{and}
  \bibinfo{author}{\bibfnamefont{M.}~\bibnamefont{Tortola}},
  \bibinfo{journal}{Front.in Phys.} \textbf{\bibinfo{volume}{6}},
  \bibinfo{pages}{10} (\bibinfo{year}{2018}), \eprint{1710.09360}.

\bibitem[{\citenamefont{Miranda and Nunokawa}(2015)}]{Miranda:2015dra}
\bibinfo{author}{\bibfnamefont{O.~G.} \bibnamefont{Miranda}} \bibnamefont{and}
  \bibinfo{author}{\bibfnamefont{H.}~\bibnamefont{Nunokawa}},
  \bibinfo{journal}{New J. Phys.} \textbf{\bibinfo{volume}{17}},
  \bibinfo{pages}{095002} (\bibinfo{year}{2015}), \eprint{1505.06254}.

\bibitem[{\citenamefont{Ohlsson}(2013)}]{Ohlsson:2012kf}
\bibinfo{author}{\bibfnamefont{T.}~\bibnamefont{Ohlsson}},
  \bibinfo{journal}{Rept. Prog. Phys.} \textbf{\bibinfo{volume}{76}},
  \bibinfo{pages}{044201} (\bibinfo{year}{2013}), \eprint{1209.2710}.

\bibitem[{\citenamefont{Bhupal~Dev et~al.}(2019)}]{Dev:2019anc}
\bibinfo{author}{\bibfnamefont{P.~S.} \bibnamefont{Bhupal~Dev}}
\bibnamefont{et~al.}
  \bibinfo{journal}{SciPost Phys. Proc.} \textbf{\bibinfo{volume}{2}},
  \bibinfo{pages}{001} (\bibinfo{year}{2019}),
\eprint{1907.00991}.

\bibitem[{\citenamefont{Barranco et~al.}(2005)\citenamefont{Barranco, Miranda,
  and Rashba}}]{Barranco:2005yy}
\bibinfo{author}{\bibfnamefont{J.}~\bibnamefont{Barranco}},
  \bibinfo{author}{\bibfnamefont{O.~G.} \bibnamefont{Miranda}},
  \bibnamefont{and} \bibinfo{author}{\bibfnamefont{T.~I.}
  \bibnamefont{Rashba}}, \bibinfo{journal}{JHEP} \textbf{\bibinfo{volume}{12}},
  \bibinfo{pages}{021} (\bibinfo{year}{2005}), \eprint{hep-ph/0508299}.

\bibitem[{\citenamefont{Esteban et~al.}(2018)\citenamefont{Esteban,
  Gonzalez-Garcia, Maltoni, Martinez-Soler, and Salvado}}]{Esteban:2018ppq}
\bibinfo{author}{\bibfnamefont{I.}~\bibnamefont{Esteban}},
  \bibinfo{author}{\bibfnamefont{M.~C.} \bibnamefont{Gonzalez-Garcia}},
  \bibinfo{author}{\bibfnamefont{M.}~\bibnamefont{Maltoni}},
  \bibinfo{author}{\bibfnamefont{I.}~\bibnamefont{Martinez-Soler}},
  \bibnamefont{and} \bibinfo{author}{\bibfnamefont{J.}~\bibnamefont{Salvado}},
  \bibinfo{journal}{JHEP} \textbf{\bibinfo{volume}{08}}, \bibinfo{pages}{180}
  (\bibinfo{year}{2018}), \eprint{1805.04530}.

\bibitem[{\citenamefont{Aristizabal~Sierra
  et~al.}(2018{\natexlab{b}})\citenamefont{Aristizabal~Sierra, De~Romeri, and
  Rojas}}]{AristizabalSierra:2018eqm}
\bibinfo{author}{\bibfnamefont{D.}~\bibnamefont{Aristizabal~Sierra}},
  \bibinfo{author}{\bibfnamefont{V.}~\bibnamefont{De~Romeri}},
  \bibnamefont{and} \bibinfo{author}{\bibfnamefont{N.}~\bibnamefont{Rojas}},
  \bibinfo{journal}{Phys. Rev.} \textbf{\bibinfo{volume}{D98}},
  \bibinfo{pages}{075018} (\bibinfo{year}{2018}{\natexlab{b}}),
  \eprint{1806.07424}.

\bibitem[{\citenamefont{de~Salas et~al.}(2017)\citenamefont{de~Salas, Forero,
  Ternes, Tortola, and Valle}}]{deSalas:2017kay}
\bibinfo{author}{\bibfnamefont{P.~F.} \bibnamefont{de~Salas}},
  \bibinfo{author}{\bibfnamefont{D.~V.} \bibnamefont{Forero}},
  \bibinfo{author}{\bibfnamefont{C.~A.} \bibnamefont{Ternes}},
  \bibinfo{author}{\bibfnamefont{M.}~\bibnamefont{Tortola}}, \bibnamefont{and}
  \bibinfo{author}{\bibfnamefont{J.~W.~F.} \bibnamefont{Valle}}
  \bibinfo{journal}{Phys. Lett.} \textbf{\bibinfo{volume}{B782}},
  \bibinfo{pages}{633} (\bibinfo{year}{2018}),
  \eprint{1708.01186}.

\bibitem[{\citenamefont{Valencia-Globalfit}(2018)}]{globalfit}
\bibinfo{author}{\bibnamefont{Valencia-Globalfit}},
  \bibinfo{howpublished}{\url{http://globalfit.astroparticles.es/}}
  (\bibinfo{year}{2018}).

\bibitem[{\citenamefont{Esteban et~al.}(2017)\citenamefont{Esteban,
  Gonzalez-Garcia, Maltoni, Martinez-Soler, and Schwetz}}]{Esteban:2016qun}
\bibinfo{author}{\bibfnamefont{I.}~\bibnamefont{Esteban}},
  \bibinfo{author}{\bibfnamefont{M.~C.} \bibnamefont{Gonzalez-Garcia}},
  \bibinfo{author}{\bibfnamefont{M.}~\bibnamefont{Maltoni}},
  \bibinfo{author}{\bibfnamefont{I.}~\bibnamefont{Martinez-Soler}},
  \bibnamefont{and} \bibinfo{author}{\bibfnamefont{T.}~\bibnamefont{Schwetz}},
  \bibinfo{journal}{JHEP} \textbf{\bibinfo{volume}{01}}, \bibinfo{pages}{087}
  (\bibinfo{year}{2017}), \eprint{1611.01514}.

\bibitem[{\citenamefont{Capozzi et~al.}(2018)\citenamefont{Capozzi, Lisi,
  Marrone, and Palazzo}}]{Capozzi:2018ubv}
\bibinfo{author}{\bibfnamefont{F.}~\bibnamefont{Capozzi}},
  \bibinfo{author}{\bibfnamefont{E.}~\bibnamefont{Lisi}},
  \bibinfo{author}{\bibfnamefont{A.}~\bibnamefont{Marrone}}, \bibnamefont{and}
  \bibinfo{author}{\bibfnamefont{A.}~\bibnamefont{Palazzo}},
  \bibinfo{journal}{Prog. Part. Nucl. Phys.} \textbf{\bibinfo{volume}{102}},
  \bibinfo{pages}{48} (\bibinfo{year}{2018}), \eprint{1804.09678}.

\bibitem[{\citenamefont{Miranda et~al.}(2006)\citenamefont{Miranda, Tortola,
  and Valle}}]{miranda:2004nb}
\bibinfo{author}{\bibfnamefont{O.~G.} \bibnamefont{Miranda}},
  \bibinfo{author}{\bibfnamefont{M.~A.} \bibnamefont{Tortola}},
  \bibnamefont{and} \bibinfo{author}{\bibfnamefont{J.~W.~F.}
  \bibnamefont{Valle}}, \bibinfo{journal}{JHEP} \textbf{\bibinfo{volume}{10}},
  \bibinfo{pages}{008} (\bibinfo{year}{2006}).

\bibitem[{\citenamefont{Coloma and Schwetz}(2016)}]{Coloma:2016gei}
\bibinfo{author}{\bibfnamefont{P.}~\bibnamefont{Coloma}} \bibnamefont{and}
  \bibinfo{author}{\bibfnamefont{T.}~\bibnamefont{Schwetz}},
  \bibinfo{journal}{Phys. Rev.} \textbf{\bibinfo{volume}{D94}},
  \bibinfo{pages}{055005} (\bibinfo{year}{2016}), \bibinfo{note}{[Erratum:
  Phys. Rev.D95,no.7,079903(2017)]}, \eprint{1604.05772}.

\bibitem[{\citenamefont{Miranda et~al.}(2016)\citenamefont{Miranda, Tortola,
  and Valle}}]{Miranda:2016wdr}
\bibinfo{author}{\bibfnamefont{O.~G.} \bibnamefont{Miranda}},
  \bibinfo{author}{\bibfnamefont{M.}~\bibnamefont{Tortola}}, \bibnamefont{and}
  \bibinfo{author}{\bibfnamefont{J.~W.~F.} \bibnamefont{Valle}},
  \bibinfo{journal}{Phys. Rev. Lett.} \textbf{\bibinfo{volume}{117}},
  \bibinfo{pages}{061804} (\bibinfo{year}{2016}), \eprint{1604.05690}.

\bibitem[{\citenamefont{Forero and Huber}(2016)}]{Forero:2016cmb}
\bibinfo{author}{\bibfnamefont{D.~V.} \bibnamefont{Forero}} \bibnamefont{and}
  \bibinfo{author}{\bibfnamefont{P.}~\bibnamefont{Huber}},
  \bibinfo{journal}{Phys. Rev. Lett.} \textbf{\bibinfo{volume}{117}},
  \bibinfo{pages}{031801} (\bibinfo{year}{2016}), \eprint{1601.03736}.

\bibitem[{\citenamefont{Drukier and Stodolsky}(1984)}]{Drukier:1983gj}
\bibinfo{author}{\bibfnamefont{A.}~\bibnamefont{Drukier}} \bibnamefont{and}
  \bibinfo{author}{\bibfnamefont{L.}~\bibnamefont{Stodolsky}},
  \bibinfo{journal}{Phys. Rev.} \textbf{\bibinfo{volume}{D30}},
  \bibinfo{pages}{2295} (\bibinfo{year}{1984}), \bibinfo{note}{[,395(1984)]}.

\bibitem[{\citenamefont{Patton et~al.}(2012)\citenamefont{Patton, Engel,
  McLaughlin, and Schunck}}]{Patton:2012jr}
\bibinfo{author}{\bibfnamefont{K.}~\bibnamefont{Patton}},
  \bibinfo{author}{\bibfnamefont{J.}~\bibnamefont{Engel}},
  \bibinfo{author}{\bibfnamefont{G.~C.} \bibnamefont{McLaughlin}},
  \bibnamefont{and} \bibinfo{author}{\bibfnamefont{N.}~\bibnamefont{Schunck}},
  \bibinfo{journal}{Phys. Rev.} \textbf{\bibinfo{volume}{C86}},
  \bibinfo{pages}{024612} (\bibinfo{year}{2012}), \eprint{1207.0693}.

\bibitem[{\citenamefont{Papoulias and Kosmas}(2015)}]{Papoulias:2015vxa}
\bibinfo{author}{\bibfnamefont{D.~K.} \bibnamefont{Papoulias}}
  \bibnamefont{and} \bibinfo{author}{\bibfnamefont{T.~S.}
  \bibnamefont{Kosmas}}, \bibinfo{journal}{Adv. High Energy Phys.}
  \textbf{\bibinfo{volume}{2015}}, \bibinfo{pages}{763648}
  (\bibinfo{year}{2015}), \eprint{1502.02928}.

\bibitem[{\citenamefont{Bednyakov and Naumov}(2018)}]{Bednyakov:2018mjd}
\bibinfo{author}{\bibfnamefont{V.~A.} \bibnamefont{Bednyakov}}
  \bibnamefont{and} \bibinfo{author}{\bibfnamefont{D.~V.}
  \bibnamefont{Naumov}}, \bibinfo{journal}{Phys. Rev.}
  \textbf{\bibinfo{volume}{D98}}, \bibinfo{pages}{053004}
  (\bibinfo{year}{2018}), \eprint{1806.08768}.

\bibitem[{\citenamefont{Patrignani et~al.}(2016)}]{Patrignani:2016xqp}
\bibinfo{author}{\bibfnamefont{C.}~\bibnamefont{Patrignani}}
  \bibnamefont{et~al.} (\bibinfo{collaboration}{Particle Data Group}),
  \bibinfo{journal}{Chin. Phys.} \textbf{\bibinfo{volume}{C40}},
  \bibinfo{pages}{100001} (\bibinfo{year}{2016}).

\bibitem[{\citenamefont{Scholberg}(2006)}]{Scholberg:2005qs}
\bibinfo{author}{\bibfnamefont{K.}~\bibnamefont{Scholberg}},
  \bibinfo{journal}{Phys. Rev.} \textbf{\bibinfo{volume}{D73}},
  \bibinfo{pages}{033005} (\bibinfo{year}{2006}), \eprint{hep-ex/0511042}.

\bibitem[{\citenamefont{Dent et~al.}(2018)\citenamefont{Dent, Dutta, Liao,
  Newstead, Strigari, and Walker}}]{Dent:2017mpr}
\bibinfo{author}{\bibfnamefont{J.~B.} \bibnamefont{Dent}},
  \bibinfo{author}{\bibfnamefont{B.}~\bibnamefont{Dutta}},
  \bibinfo{author}{\bibfnamefont{S.}~\bibnamefont{Liao}},
  \bibinfo{author}{\bibfnamefont{J.~L.} \bibnamefont{Newstead}},
  \bibinfo{author}{\bibfnamefont{L.~E.} \bibnamefont{Strigari}},
  \bibnamefont{and} \bibinfo{author}{\bibfnamefont{J.~W.}
  \bibnamefont{Walker}}, \bibinfo{journal}{Phys. Rev.}
  \textbf{\bibinfo{volume}{D97}}, \bibinfo{pages}{035009}
  (\bibinfo{year}{2018}), \eprint{1711.03521}.

\bibitem[{\citenamefont{Coloma et~al.}(2017{\natexlab{a}})\citenamefont{Coloma,
  Denton, Gonzalez-Garcia, Maltoni, and Schwetz}}]{Coloma:2017egw}
\bibinfo{author}{\bibfnamefont{P.}~\bibnamefont{Coloma}},
  \bibinfo{author}{\bibfnamefont{P.~B.} \bibnamefont{Denton}},
  \bibinfo{author}{\bibfnamefont{M.~C.} \bibnamefont{Gonzalez-Garcia}},
  \bibinfo{author}{\bibfnamefont{M.}~\bibnamefont{Maltoni}}, \bibnamefont{and}
  \bibinfo{author}{\bibfnamefont{T.}~\bibnamefont{Schwetz}},
  \bibinfo{journal}{JHEP} \textbf{\bibinfo{volume}{04}}, \bibinfo{pages}{116}
  (\bibinfo{year}{2017}{\natexlab{a}}), \eprint{1701.04828}.

\bibitem[{\citenamefont{Papoulias
  et~al.}(2019{\natexlab{b}})\citenamefont{Papoulias, Kosmas, Sahu, Kota, and
  Hota}}]{Papoulias:2019lfi}
\bibinfo{author}{\bibfnamefont{D.~K.} \bibnamefont{Papoulias}},
  \bibinfo{author}{\bibfnamefont{T.~S.} \bibnamefont{Kosmas}},
  \bibinfo{author}{\bibfnamefont{R.}~\bibnamefont{Sahu}},
  \bibinfo{author}{\bibfnamefont{V.~K.~B.} \bibnamefont{Kota}},
  \bibnamefont{and} \bibinfo{author}{\bibfnamefont{M.}~\bibnamefont{Hota}}
  \bibinfo{journal}{Phys. Lett.} \textbf{\bibinfo{volume}{B800}},
  \bibinfo{pages}{135133} (\bibinfo{year}{2020}),
  \eprint{1903.03722}.

\bibitem[{\citenamefont{Coloma et~al.}(2017{\natexlab{b}})\citenamefont{Coloma,
  Gonzalez-Garcia, Maltoni, and Schwetz}}]{Coloma:2017ncl}
\bibinfo{author}{\bibfnamefont{P.}~\bibnamefont{Coloma}},
  \bibinfo{author}{\bibfnamefont{M.~C.} \bibnamefont{Gonzalez-Garcia}},
  \bibinfo{author}{\bibfnamefont{M.}~\bibnamefont{Maltoni}}, \bibnamefont{and}
  \bibinfo{author}{\bibfnamefont{T.}~\bibnamefont{Schwetz}},
  \bibinfo{journal}{Phys. Rev.} \textbf{\bibinfo{volume}{D96}},
  \bibinfo{pages}{115007} (\bibinfo{year}{2017}{\natexlab{b}}),
  \eprint{1708.02899}.

\bibitem[{\citenamefont{Miranda
  et~al.}(2019{\natexlab{b}})\citenamefont{Miranda, Papoulias, Tórtola, and
  Valle}}]{Miranda:2019wdy}
\bibinfo{author}{\bibfnamefont{O.~G.} \bibnamefont{Miranda}},
  \bibinfo{author}{\bibfnamefont{D.~K.} \bibnamefont{Papoulias}},
  \bibinfo{author}{\bibfnamefont{M.}~\bibnamefont{Tórtola}}, \bibnamefont{and}
  \bibinfo{author}{\bibfnamefont{J.~W.~F.} \bibnamefont{Valle}},
  \bibinfo{journal}{JHEP} \textbf{\bibinfo{volume}{07}}, \bibinfo{pages}{103}
  (\bibinfo{year}{2019}{\natexlab{b}}), \eprint{1905.03750}.

\bibitem[{\citenamefont{Parada}(2019)}]{Parada:2019gvy}
\bibinfo{author}{\bibfnamefont{A.}~\bibnamefont{Parada}}
  (\bibinfo{year}{2019}), \eprint{1907.04942}.

\bibitem[{\citenamefont{Reines and Cowan}(1956)}]{Reines:1956rs}
\bibinfo{author}{\bibfnamefont{F.}~\bibnamefont{Reines}} \bibnamefont{and}
  \bibinfo{author}{\bibfnamefont{C.~L.} \bibnamefont{Cowan}},
  \bibinfo{journal}{Nature} \textbf{\bibinfo{volume}{178}},
  \bibinfo{pages}{446} (\bibinfo{year}{1956}).

\bibitem[{\citenamefont{Cowan et~al.}(1956)\citenamefont{Cowan, Reines,
  Harrison, Kruse, and McGuire}}]{Cowan:1992xc}
\bibinfo{author}{\bibfnamefont{C.~L.} \bibnamefont{Cowan}},
  \bibinfo{author}{\bibfnamefont{F.}~\bibnamefont{Reines}},
  \bibinfo{author}{\bibfnamefont{F.~B.} \bibnamefont{Harrison}},
  \bibinfo{author}{\bibfnamefont{H.~W.} \bibnamefont{Kruse}}, \bibnamefont{and}
  \bibinfo{author}{\bibfnamefont{A.~D.} \bibnamefont{McGuire}},
  \bibinfo{journal}{Science} \textbf{\bibinfo{volume}{124}},
  \bibinfo{pages}{103} (\bibinfo{year}{1956}).

\bibitem[{\citenamefont{An et~al.}(2012)}]{An:2012eh}
\bibinfo{author}{\bibfnamefont{F.~P.} \bibnamefont{An}} \bibnamefont{et~al.}
  (\bibinfo{collaboration}{Daya Bay}), \bibinfo{journal}{Phys. Rev. Lett.}
  \textbf{\bibinfo{volume}{108}}, \bibinfo{pages}{171803}
  (\bibinfo{year}{2012}), \eprint{1203.1669}.

\bibitem[{\citenamefont{Ahn et~al.}(2012)}]{Ahn:2012nd}
\bibinfo{author}{\bibfnamefont{J.~K.} \bibnamefont{Ahn}} \bibnamefont{et~al.}
  (\bibinfo{collaboration}{RENO}), \bibinfo{journal}{Phys. Rev. Lett.}
  \textbf{\bibinfo{volume}{108}}, \bibinfo{pages}{191802}
  (\bibinfo{year}{2012}), \eprint{1204.0626}.

\bibitem[{\citenamefont{Abe et~al.}(2012)}]{Abe:2011fz}
\bibinfo{author}{\bibfnamefont{Y.}~\bibnamefont{Abe}} \bibnamefont{et~al.}
  (\bibinfo{collaboration}{Double Chooz}), \bibinfo{journal}{Phys. Rev. Lett.}
  \textbf{\bibinfo{volume}{108}}, \bibinfo{pages}{131801}
  (\bibinfo{year}{2012}), \eprint{1112.6353}.

\bibitem[{\citenamefont{Apollonio et~al.}(1999)}]{Apollonio:1999ae}
\bibinfo{author}{\bibfnamefont{M.}~\bibnamefont{Apollonio}}
  \bibnamefont{et~al.} (\bibinfo{collaboration}{CHOOZ}),
  \bibinfo{journal}{Phys. Lett.} \textbf{\bibinfo{volume}{B466}},
  \bibinfo{pages}{415} (\bibinfo{year}{1999}), \eprint{hep-ex/9907037}.

\bibitem[{\citenamefont{Deniz et~al.}(2010)}]{Deniz:2009mu}
\bibinfo{author}{\bibfnamefont{M.}~\bibnamefont{Deniz}} \bibnamefont{et~al.}
  (\bibinfo{collaboration}{TEXONO collaboration}), \bibinfo{journal}{Phys.
  Rev.} \textbf{\bibinfo{volume}{D81}}, \bibinfo{pages}{072001}
  (\bibinfo{year}{2010}), \eprint{0911.1597}.

\bibitem[{\citenamefont{Reines et~al.}(1976)\citenamefont{Reines, Gurr, and
  Sobel}}]{Reines:1976pv}
\bibinfo{author}{\bibfnamefont{F.}~\bibnamefont{Reines}},
  \bibinfo{author}{\bibfnamefont{H.~S.} \bibnamefont{Gurr}}, \bibnamefont{and}
  \bibinfo{author}{\bibfnamefont{H.~W.} \bibnamefont{Sobel}},
  \bibinfo{journal}{Phys. Rev. Lett.} \textbf{\bibinfo{volume}{37}},
  \bibinfo{pages}{315} (\bibinfo{year}{1976}).

\bibitem[{\citenamefont{Mueller et~al.}(2011)\citenamefont{Mueller, Lhuillier,
  Fallot, Letourneau, Cormon et~al.}}]{Mueller:2011nm}
\bibinfo{author}{\bibfnamefont{T.}~\bibnamefont{Mueller}},
  \bibinfo{author}{\bibfnamefont{D.}~\bibnamefont{Lhuillier}},
  \bibinfo{author}{\bibfnamefont{M.}~\bibnamefont{Fallot}},
  \bibinfo{author}{\bibfnamefont{A.}~\bibnamefont{Letourneau}},
  \bibinfo{author}{\bibfnamefont{S.}~\bibnamefont{Cormon}},
  \bibnamefont{et~al.}, \bibinfo{journal}{Phys.Rev.}
  \textbf{\bibinfo{volume}{C83}}, \bibinfo{pages}{054615}
  (\bibinfo{year}{2011}), \eprint{1101.2663}.

\bibitem[{\citenamefont{Huber}(2011)}]{Huber:2011wv}
\bibinfo{author}{\bibfnamefont{P.}~\bibnamefont{Huber}},
  \bibinfo{journal}{Phys. Rev.} \textbf{\bibinfo{volume}{C84}},
  \bibinfo{pages}{024617} (\bibinfo{year}{2011}), \bibinfo{note}{[Erratum:
  Phys. Rev.C85,029901(2012)]}, \eprint{1106.0687}.

\bibitem[{\citenamefont{Kopeikin et~al.}(1997)\citenamefont{Kopeikin,
  Mikaelyan, and Sinev}}]{Kopeikin:1997ve}
\bibinfo{author}{\bibfnamefont{V.~I.} \bibnamefont{Kopeikin}},
  \bibinfo{author}{\bibfnamefont{L.~A.} \bibnamefont{Mikaelyan}},
  \bibnamefont{and} \bibinfo{author}{\bibfnamefont{V.~V.} \bibnamefont{Sinev}},
  \bibinfo{journal}{Phys. Atom. Nucl.} \textbf{\bibinfo{volume}{60}},
  \bibinfo{pages}{172} (\bibinfo{year}{1997}), \bibinfo{note}{[Yad.
  Fiz.60,230(1997)]}.

\bibitem[{\citenamefont{Huber and Schwetz}(2004)}]{Huber:2004xh}
\bibinfo{author}{\bibfnamefont{P.}~\bibnamefont{Huber}} \bibnamefont{and}
  \bibinfo{author}{\bibfnamefont{T.}~\bibnamefont{Schwetz}},
  \bibinfo{journal}{Phys. Rev.} \textbf{\bibinfo{volume}{D70}},
  \bibinfo{pages}{053011} (\bibinfo{year}{2004}), \eprint{hep-ph/0407026}.

\bibitem[{\citenamefont{Akimov et~al.}(2018{\natexlab{a}})}]{Akimov:2018ghi}
\bibinfo{author}{\bibfnamefont{D.}~\bibnamefont{Akimov}} \bibnamefont{et~al.}
  (\bibinfo{collaboration}{COHERENT}) (\bibinfo{year}{2018}{\natexlab{a}}),
  \eprint{1803.09183}.

\bibitem[{\citenamefont{Piekarewicz et~al.}(2016)\citenamefont{Piekarewicz,
  Linero, Giuliani, and Chicken}}]{Piekarewicz:2016vbn}
\bibinfo{author}{\bibfnamefont{J.}~\bibnamefont{Piekarewicz}},
  \bibinfo{author}{\bibfnamefont{A.~R.} \bibnamefont{Linero}},
  \bibinfo{author}{\bibfnamefont{P.}~\bibnamefont{Giuliani}}, \bibnamefont{and}
  \bibinfo{author}{\bibfnamefont{E.}~\bibnamefont{Chicken}},
  \bibinfo{journal}{Phys. Rev.} \textbf{\bibinfo{volume}{C94}},
  \bibinfo{pages}{034316} (\bibinfo{year}{2016}), \eprint{1604.07799}.

\bibitem[{\citenamefont{Kolbe and Langanke}(2001)}]{Kolbe:2000np}
\bibinfo{author}{\bibfnamefont{E.}~\bibnamefont{Kolbe}} \bibnamefont{and}
  \bibinfo{author}{\bibfnamefont{K.}~\bibnamefont{Langanke}},
  \bibinfo{journal}{Phys. Rev.} \textbf{\bibinfo{volume}{C63}},
  \bibinfo{pages}{025802} (\bibinfo{year}{2001}), \eprint{nucl-th/0003060}.

\bibitem[{\citenamefont{Akimov et~al.}(2018{\natexlab{b}})}]{Akimov:2018vzs}
\bibinfo{author}{\bibfnamefont{D.}~\bibnamefont{Akimov}} \bibnamefont{et~al.}
  (\bibinfo{collaboration}{COHERENT}) (\bibinfo{year}{2018}{\natexlab{b}}),
  \eprint{1804.09459}.

\bibitem[{\citenamefont{Collar et~al.}(2019)\citenamefont{Collar, Kavner, and
  Lewis}}]{Collar:2019ihs}
\bibinfo{author}{\bibfnamefont{J.~I.} \bibnamefont{Collar}},
  \bibinfo{author}{\bibfnamefont{A.~R.~L.} \bibnamefont{Kavner}},
  \bibnamefont{and} \bibinfo{author}{\bibfnamefont{C.~M.} \bibnamefont{Lewis}},
  \bibinfo{journal}{Phys. Rev.} \textbf{\bibinfo{volume}{D100}},
  \bibinfo{pages}{033003} (\bibinfo{year}{2019}), \eprint{1907.04828}.

\bibitem[{\citenamefont{Baxter et~al.}(2019)}]{Baxter:2019mcx}
\bibinfo{author}{\bibfnamefont{D.}~\bibnamefont{Baxter}} \bibnamefont{et~al.}
  (\bibinfo{year}{2019}), \eprint{1911.00762}.

\end{thebibliography}
\end{document}